\documentclass[aps,pre,reprint,groupedaddress,showpacs,amsmath]{revtex4-1}
\usepackage{mathtools}
\usepackage{amssymb}
\usepackage{graphicx}
\usepackage[colorlinks=true,linkcolor=blue,urlcolor=blue,citecolor=blue,anchorcolor=blue]{hyperref}
\usepackage{multirow}
\usepackage{color}

\newcommand{\modified}[1]{#1} 
\newcommand{\modljd}[1]{#1}

\newcommand{\pr}[1]{\left(#1 \right)} 
\newcommand{\br}[1]{\left[#1 \right]} 
\newcommand{\cbrace}[1]{\left\{#1 \right\}} 

\newcommand{\avg}[1]{\left< #1 \right>} 
\renewcommand{\d}[2]{\frac{\mathrm{d} #1}{\mathrm{d} #2}} 

\let\basetop=\top
\renewcommand{\top}{{}^\basetop \!}
\newcommand{\dx}{\: \mathrm{d}}
\newcommand{\fa}{\: \forall \:}

\renewcommand{\implies}{\, \Rightarrow \,}

\begin{document}

\title{Phase transition of the susceptible-infected-susceptible dynamics on time-varying configuration model networks}
\author{Guillaume St-Onge}
\email[]{guillaume.st-onge.4@ulaval.ca}
\author{Jean-Gabriel Young}
\author{Edward Laurence}
\author{Charles Murphy}
\author{Louis J. \surname{Dub\'e}}
\email[]{Louis.Dube@phy.ulaval.ca}
\affiliation{D\'epartement de Physique, de G\'enie Physique, et d'Optique, 
Universit\'e Laval, Qu\'ebec (Qu\'ebec), Canada, G1V 0A6}

\date{\today}

\begin{abstract}
We present a degree-based theoretical framework to study the susceptible-infected-susceptible (SIS) dynamics on time-varying (rewired) configuration model networks. 
\modljd{Using this framework on a given degree distribution, we provide a detailed analysis of the stationary state using the rewiring rate to explore the whole range of 
the time variation of the structure relative to that of the SIS process.}
 This analysis is suitable for the characterization of the phase transition and leads to three main contributions. (i) We obtain a self-consistent expression for the absorbing-state threshold, able to capture both collective and hub activation. (ii) We recover the predictions of a number of existing approaches as limiting cases of our analysis, providing thereby a unifying point of view for the SIS dynamics on random networks. (iii) \modified{We obtain bounds for the critical exponents of a number of quantities in the stationary state.} This allows us to reinterpret the concept of hub-dominated phase transition. Within our framework, it appears as a heterogeneous critical phenomenon : observables for different degree classes have a different scaling with the infection rate. \modified{This phenomenon is followed by the successive activation of the degree classes beyond the epidemic threshold.}
\end{abstract}


\maketitle



\section{Introduction}

\modified{The susceptible-infected-susceptible (SIS) model is one of the classical and most studied models of disease propagation on complex networks} \cite{Barrat2008,Newman2010,Pastor-Satorras2015}. It can be understood as a specific case of binary-state dynamics \cite{Gleeson2011,Gleeson2013} where nodes are either susceptible $(S)$ or infected $(I)$. Susceptible nodes become infected at rate $\lambda l$ where $l$ represents the number of infected neighbors; infected nodes recover and become susceptible at rate $\mu$, set to unity without loss of generality. 
\modified{ Despite being a crude approximation of reality, this is arguably one of the simplest models leading to an absorbing-state phase transition. For infinite size networks in the stationary state ($t \to \infty$), there are two distinct phases~:} 
an \emph{absorbing phase}---consisting of all nodes being susceptible---and an \emph{active phase} where a constant fraction of the nodes remains infected on average. The former is attractive for any initial configurations with infection rate $\lambda \leq \lambda_c$, which defines the threshold $\lambda_c$. From a statistical physics perspective, this represents a critical phenomenon, where the density of infected nodes in the stationary state plays the role of the order parameter.

It is now common knowledge in network science that the degree distribution $P(k)$, the probability that a random node has $k$ neighbors, is a fundamental property to quantify the extent of an epidemic outbreak \cite{Barrat2008,Pastor-Satorras2015}. To this end, random networks with an arbitrary degree distribution have been extensively used to study the impact of this property on the spreading of diseases \cite{Boguna2002,Castellano2010,Gleeson2011,Castellano2012,Ferreira2012,Mata2013,Mata2014,Mata2015,Pastor-Satorras2015,Cai2016,Cota2016,Ferreira2016}. Recently, a profound impact of the degree distribution has been unveiled, leading to an interesting dichotomy for the nature of the phase transition of the SIS model on networks. The activity just beyond the threshold is either localized in the neighborhood of high degree nodes (hubs), sustained by correlated reinfections, or maintained collectively by the whole network \cite{Castellano2010,Castellano2012,Mata2015,Ferreira2016}. As in Ref.~\cite{Ferreira2016}, we will use the terminology \textit{hub activation} and \textit{collective activation} to discriminate these two scenarios.

To capture the dynamics and describe its critical behavior, various analytical approaches have been developed using mean field, pair approximation and dynamic message passing techniques \cite{Boguna2002,VanMieghem2009,Gleeson2011,Cator2012,Mata2013,Mata2014,Shrestha2015,Cai2016} (see Refs.~\cite{Pastor-Satorras2015,Wang2017} for recent reviews). They can be divided into two major families : \emph{degree}-based and \emph{individual}-based formalisms. The former is a compartmental modeling scheme that assumes the statistical equivalence of each node in a same degree class. It leads to simple approaches with explicit analytical predictions, but restricted to infinite size random networks. The latter relies explicitly on the (quenched) structure, described by an adjacency matrix $a_{ij}$, to estimate the marginal probability of infection for each node. \modljd{Its range of applicability is not restricted to infinite size random networks,} \modified{but it is less amenable to analytical treatment than degree-based approaches.}

Despite the same basic structural information---the degree distribution---there remain disparities between the predictions of degree-based and individual-based formalisms. An important theoretical gap that needs to be addressed is that current characterizations of the phase transition using degree-based approaches are unable to describe a hub activation correctly. \modljd{This arises from the fact that the neighborhood of nodes for each degree class is not described properly.}

We provide in the following a degree-based theoretical analysis of the SIS dynamics on time-varying (edges are being rewired) random networks with a fixed degree sequence in the infinite size limit. Our emphasis is on the characterization of the critical phenomenon for both, collective and hub activation. \modified{Our \emph{rewired network approach} (RNA) permits us to simulate an effective structural dynamics and mathematically provides an interpolation between existing compartmental formalisms.}

The paper is organized as follows. In Sec.~\ref{sec:mathematical_framework}, we introduce a compartmental formalism to characterize the dynamics and we show how it is related to other approaches. In Sec.~\ref{sec:stationarySolution}, we obtain the stationary distributions that we develop near the absorbing phase. Using this framework, we draw a general portrait of the phase transition. In Sec.~\ref{sec:threshold}, we present an explicit upper bound and an implicit expression for the threshold $\lambda_c$, that we compare analytically and numerically with the predictions of a number of existing approaches. In Sec.~\ref{sec:critical_exponent}, we obtain bounds for the critical exponents describing the stationary distributions near the absorbing phase, bringing to light a heterogeneous critical phenomenon associated with the hub activation. In Sec.~\ref{sec:beyond}, we discuss the impacts of structural dynamics on the hub-dominated property of a phase transition, and show the successive activation of the degree classes beyond the threshold. We finally gather concluding remarks and open challenges in Sec.~\ref{sec:conclusion}. They are followed by two Appendices, giving details of the Monte-Carlo simulations (Appendix \ref{app:monte_carlo}) and of the mathematical developments for the critical exponents (Appendix \ref{app:SM_critical_exponent}).

\section{Mathematical framework \label{sec:mathematical_framework}}

\modified{Time variations of the structure greatly affect the propagation \cite{Gross2006,Marceau2010,Holme2012,Vazquez2007,Perra2012,Taylor2012,Valdano2015}. For networks whose evolution is independent from the dynamical state \cite{Perra2012,Taylor2012,Valdano2015}, it has been shown to notably alter the epidemic threshold of the SIS model. For \emph{adaptive} networks \cite{Gross2009} where the dynamical state influences the evolution of the structure, a hysteresis loop and a first order transition have even been observed \cite{Gross2006,Marceau2010}.}

\begin{figure}
\centering
\includegraphics[width = 0.4\textwidth]{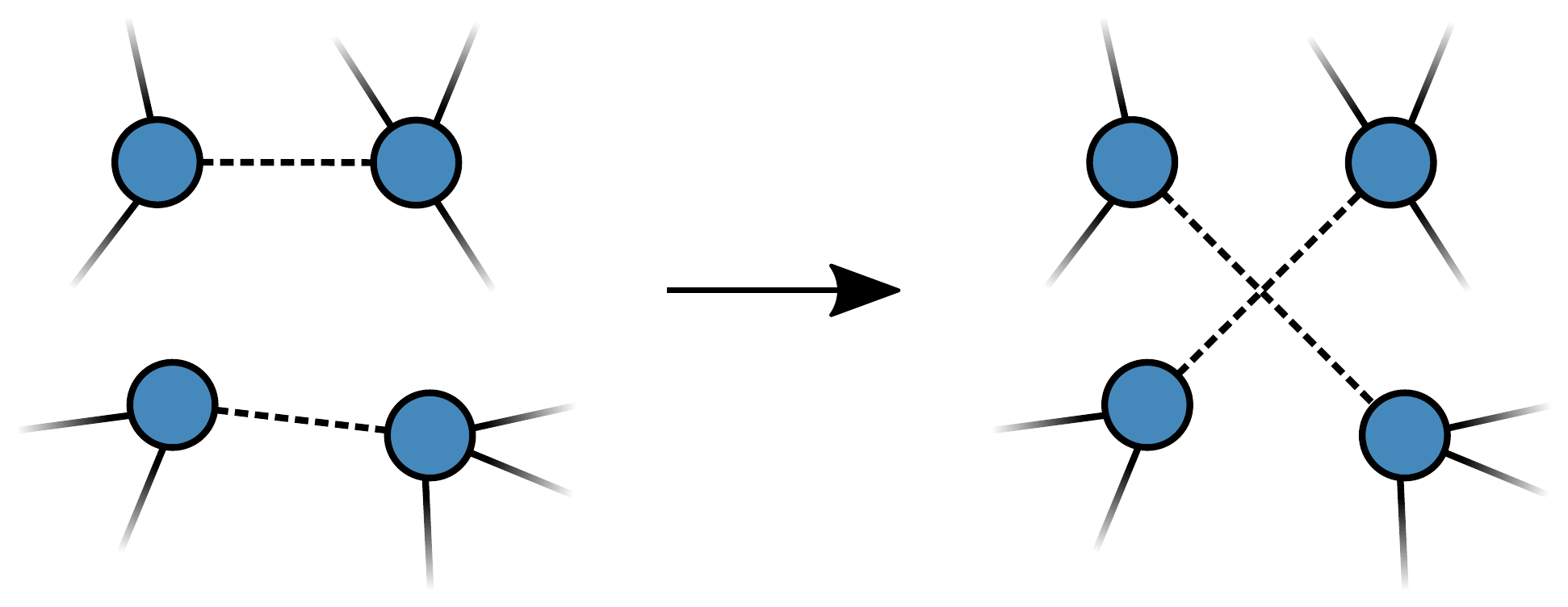}%
\caption{One move of an edge switch to eventually sample the whole of the  configuration model space. \label{fig:rewiring_mechanism}}
\end{figure}

\modified{In this paper, we consider the former scenario, a structure evolving according to a continuous Markov process, independent of the SIS dynamics. Each edge in the network is rewired at a constant rate $\omega$: a rewiring event involves two edges that are disconnected, and the stubs are rematched as presented in Fig.~\ref{fig:rewiring_mechanism}. For nodes, this implies that their stubs are effectively reconnected to random stubs in the network at the rate $\omega$. We allow loops and multiple edges to simplify the rewiring procedure and impose a structural cut-off for the maximal degree $k_\mathrm{max} < N^{1/2}$ to have a vanishing fraction of these undesired edges.} 

\modified{
This process samples a configuration model ensemble by leaving the degree sequence unaltered \cite{Fosdick2016}. Noteworthy, this allows us to control the heterogeneity of the structure independently from the time-varying mechanism. Moreover, the networks ensemble is \emph{uncorrelated}, i.e the degrees at the end points of any edge are independent.}

\modljd{
Since the structural dynamics is a Poisson process, exponentially distributed lifetimes for the edges are produced. Although it has been argued that many real contact patterns are better represented by power-law distributed lifetimes \cite{Vazquez2007,Holme2012}, our framework still captures the essence of a time-varying structure and is simple enough to lend itself to explicit analytical results. For all ensuing mathematical developments, the thermodynamic limit ($N \to \infty$) is assumed.}

\subsection{Compartmental formalism}

Since we consider a time-varying network preserving the degree sequence, the statistical equivalence of each node with a same degree $k$ is guaranteed. This implies that the probability $\rho_k(t)$ that a node of degree $k$ is infected follows the rate equation 
\begin{align}\label{eqRhoClosed}
	\d{\rho_k}{t} &= - \rho_k + \lambda k (1- \rho_k) \theta_k \;,
\end{align}
where $\theta_k(t)$ is the probability of reaching an infected node following a random edge starting from a degree $k$ susceptible node. In the stationary limit ($\dot{\rho}_k = 0 \; \fa k$), the following relations
\begin{align}\label{rhoSS}
	\rho_k^* = \frac{\lambda k \theta_k^* }{1 + \lambda k \theta_k^* } \quad \text{or} \quad \lambda k \theta_k^* = \frac{\rho_k^*}{1- \rho_k^*} \;,
\end{align}
are obtained. Stationary values will be marked hereafter with an asterisk (*). Equation \eqref{rhoSS} expresses that a node's probability of being infected is directly related to its neighborhood's state, quantified by $\theta_k^*$. Our objective is therefore to find the most precise explicit expression for this probability, taking into account the rewiring process. In the general case, we must have a degree dependent solution to represent $\theta_k^*$.

\modified{Accordingly, we consider a \emph{pair approximation} framework as introduced in Ref.~\cite{Gleeson2011,Gleeson2013}. To include the rewiring process, we account for the probability $\Theta(t)$ that a newly rewired stub reaches an infected node }
\begin{align}\label{defBigTheta}
\Theta \equiv \frac{\avg{k \rho_k}}{\avg{k}} \;,
\end{align}
where all averages $\avg{\cdots}$ are taken over $P(k)$. Let $\phi_k(t)$ be the probability of reaching an infected node following a random edge starting from a degree $k$ \textit{infected} node. We obtain (see Appendix \ref{app:PA_derivation})
\begin{subequations}\label{eqThetaPhi_PA}
\begin{align}
	\d{\theta_k}{t} =& -\lambda \br{\theta_k + (k-1)\theta_k^2}  + r_k \phi_k + (\Omega^S + \omega \Theta) (1- \theta_k) \nonumber\\
	&-\br{1+\omega(1- \Theta)}\theta_k  - \theta_k \pr{ r_k - \lambda k \theta_k} \;, \label{eqTheta_PA}\\
	\d{\phi_k}{t} =& \lambda r_k^{-1} \br{\theta_k + (k-1)\theta_k^2}  - \phi_k + (\Omega^I + \omega \Theta) (1- \phi_k) \nonumber\\
	&- \br{1+\omega(1- \Theta)}\phi_k  +  \phi_k  \pr{ 1 - \lambda k \theta_k r_k^{-1}} \;, \label{eqPhi_PA}
\end{align}
\end{subequations}
with $r_k \equiv \rho_k/(1-\rho_k)$. Also, $\Omega^S(t)$ and $\Omega^I(t)$ are the mean infection rates for the neighbors of susceptible and infected nodes. These rates are estimated by
\begin{subequations}\label{defOmega2}
\begin{align}
	{\Omega^S} &= \lambda \frac{\avg{(1- \rho_k)(\theta_k-{\theta_k}^2)(k-1)k}}{\avg{(1- \rho_k)(1- \theta_k)k }} \;, \\
	{\Omega^I} &= \lambda \frac{\avg{(1- \rho_k)[\theta_kk + {\theta_k}^2 k(k-1)]}}{\avg{(1- \rho_k)\theta_kk}} \;.
\end{align}
\end{subequations}

\modljd{
Before going any further with the analysis, it useful to discuss the approximations involved in Eqs.~\eqref{eqThetaPhi_PA}. }
\modified{
\begin{enumerate}
	\item The mean infection rates for the neighbors ($\Omega^S$ and $\Omega^I$) are independent of the degree and are estimated from mean values over the network. An infinite size configuration model network is assumed.
	\item The pair approximation considers that, for a degree $k$ susceptible node, each neighbor is infected with an independent probability $\theta_k$.
\end{enumerate}
Compartmental formalisms based only on the first approximation (effective degree or approximated master equations \cite{Lindquist2011,Gleeson2011}) lead to excellent agreement with the corresponding stochastic processes on random networks (see Refs.~\cite{Lindquist2011,Kiss2017}). The second approximation enables us to perform a thorough stationary state analysis in the following sections. Such pairwise approximations have been shown to predict an epidemic threshold that is slightly off, but still show very good agreement with numerical simulations in contrast to mean-field theories \cite{Eames2002,Gleeson2011}.
}

\subsection{Reduction and relation to other formalisms}

The rewiring rate $\omega \geq 0$ permits us to tune the interplay between the disease propagation and the structural dynamics, for which we can distinguish two extreme limits. There is the \emph{annealed} network limit when the rewiring is much faster than the propagation dynamics ($\omega\to \infty$). It is equivalent to consider the SIS dynamics on an annealed network with adjacency matrix $a_{ij} = k_i k_j/(N\avg{k})$ \cite{Pastor-Satorras2015}. In this limit, our compartmental approach is identical to the heterogeneous mean field theory (HMF) \cite{Pastor-Satorras2001a,Pastor-Satorras2001,Boguna2002}.

For annealed networks, the \textit{dynamic correlation} and the \textit{neighborhood heterogeneity} can be neglected. On the one hand, the absence of a dynamic correlation implies that the states of neighbor nodes are independent \cite{Gleeson2012,Pastor-Satorras2015,Wang2017}. On the other hand, the absence of neighborhood heterogeneity implies that the degree of a node, on average, does not affect the state of its neighbors. From a degree-based perspective, this would mean that $\theta_k^*$ is a probability independent of the degree class. 

In contrast with the annealed limit, there is the \emph{quasi-static} network limit ($\omega \to 0$), where both the dynamic correlation and the neighborhood heterogeneity cannot be neglected. Between each rewiring event, the SIS dynamics has enough time to relax and reach a stationary distribution---temporal averages for the dynamics are then equivalent to ensemble averages on every static realization of the configuration model. In this limit, our compartmental approach is equivalent to the heterogeneous pair approximation (HPA) of Ref.~\cite{Gleeson2011}, which considers both the dynamic correlation and the neighborhood heterogeneity.

We stress that our mathematical framework (as well as HPA) is different from other pair approximation formalisms that neglect the neighborhood heterogeneity, such as the pair heterogeneous mean field theory (PHMF) \cite{Mata2014} or similar approaches \cite{Cai2016}. In the quasi-static limit, we also expect our compartmental formalism to be in agreement with individual-based approaches such as quenched mean-field theory (QMF) \cite{VanMieghem2009,VanMieghem2012,Pastor-Satorras2015} and pair QMF (PQMF) \cite{Mata2013,Cator2012}. 

The RNA effectively interpolate between HPA and HMF through the tuning of the rewiring rate $\omega$. The specific properties of each formalism are compiled in Table \ref{tab:formalism_comparison}.

\begin{table}
\caption{Comparison of the properties of various formalisms. \label{tab:formalism_comparison}}
\begin{ruledtabular}
\begin{tabular}{l c c c c}
Formalism & Individual & Degree & Dynamic & Neighborhood \\
		  & -based  &  -based  & correlation & heterogeneity \\
\hline 
HMF 	  & 	& \checkmark & & \\
PHMF 	  &  	& \checkmark & \checkmark & \\
HPA 	  &  	& \checkmark & \checkmark & \checkmark \\
QMF 	  & \checkmark & & & \checkmark \\
PQMF      & \checkmark & & \checkmark & \checkmark\\
  & & & & \\
RNA & & \checkmark & \checkmark & \checkmark \\
\end{tabular}
\end{ruledtabular}
\end{table}
 
\section{Stationary distributions \label{sec:stationarySolution}}

Solving Eqs.~\eqref{eqThetaPhi_PA} in the stationary limit for $\theta_k^*$, we find
\begin{align}\label{thetaSol}
	\theta_k^*(\omega,\lambda) &= 	\begin{dcases}
					\frac{\beta}{\kappa - 1} &\text{if } k = 1 \;,\\
					\frac{k - \kappa + \sqrt{(k - \kappa)^2 + 4 \alpha \beta(k-1)}}{2 \alpha	(k-1)} &\text{if } k > 1 \;,
				\end{dcases}
\end{align}
where the parameters are
\begin{subequations} \label{parameters}
\begin{align}
	\alpha &= \frac{1 + \omega + {\Omega^I}^* }{{\Omega^I}^*+ \omega \Theta^*} \;, \\
	\beta  &= \frac{({\Omega^S}^* + \omega \Theta^*)(2 + \omega + {\Omega^I}^*)}{\lambda ({\Omega^I}^* + \omega \Theta^*)} \;, \label{beta_def}\\
	 \kappa &= \frac{(\lambda + 1 + {\Omega^S}^* + \omega)(2+ \omega + {\Omega^I}^*) - \lambda}{\lambda({\Omega^I}^* + \omega \Theta^*)} \;.
\end{align}
\end{subequations}
As desired, we have obtained a degree dependent solution for $\theta_k^*$. At this point, one can already verify the consistency with HMF in the annealed limit~: Taking $\omega \to \infty$ in Eq.~\eqref{thetaSol}, one recovers $\theta_k^* \to \Theta^*$. 
\modified{For finite $\omega$ however, we obtain a solution that is potentially heterogeneous among degree classes.}

\subsection{Collective and hub activations \label{subsec:collective_hub}}
\modified{
As briefly discussed in the Introduction, there exists a dichotomy in the nature of the phase transition of the SIS model. Numerical evidences suggest that near the absorbing phase, the activity is localized either on the hubs (hub activation) or on the innermost network core (collective activation) \cite{Castellano2012}. This dichotomy is also supported theoretically by individual-based approaches such as QMF \cite{Goltsev2012}, for which the active phase near the epidemic threshold is dominated by the principal eigenvector of the adjacency matrix. 
\modljd{This eigenvector is localized} either on the subgraph associated with the highest degree nodes or on the shell with the largest index in the $K$-core decomposition \cite{Pastor2016,Castellano2017}.}

\modified{
For uncorrelated configuration model networks with power-law degree distribution $P(k) \sim k^{-\gamma}$, this dichotomy is reflected as two distinct regimes \cite{Goltsev2012,Castellano2012}. For $\gamma < 5/2$, the phase transition is collective due to the presence of a large innercore whereas for $\gamma \geq 5/2$, the phase transition is dominated instead by the hubs. It is important to note that these two regimes are well defined only in the thermodynamic limit ($N \to \infty$ and consequently $k_\mathrm{max} \to \infty$) \cite{Goltsev2012}.}

\begin{figure}
\includegraphics[width = 0.5\textwidth]{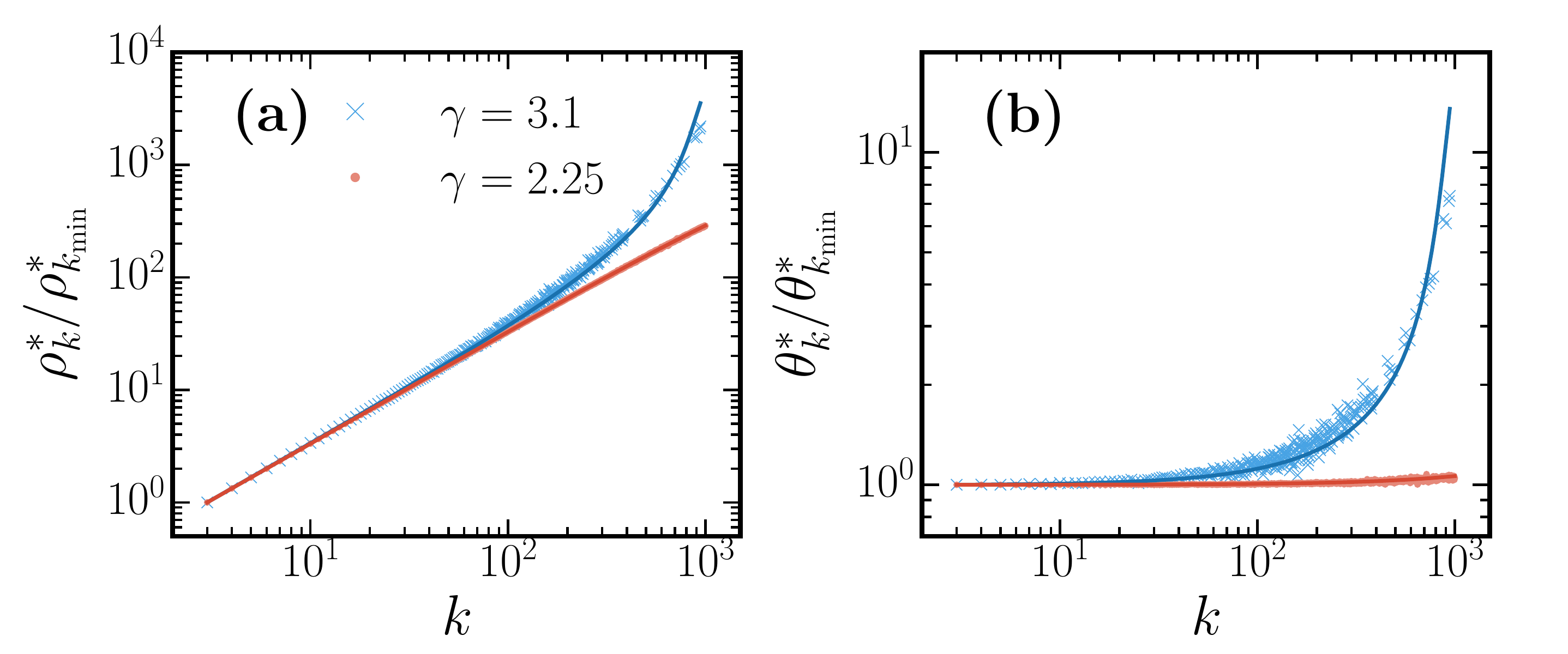}
\caption{(Color online). \modified{Degree dependent observables for the SIS model in the quasi-static limit ($\omega \to 0$), near the absorbing phase. Solid lines are the predictions of Eqs.\eqref{thetaSol} and \eqref{rhoSS} ; markers are the results of Monte-Carlo simulations. To prevent the system from reaching the absorbing state, we have sampled the quasi-stationary distribution of the system \cite{Oliveira2005,Ferreira2011,Sander2016}. Averages are made on $10^2$ realizations of the configuration model with the same degree sequence to simulate the quasi-static limit. The degree sequences of $N = 10^6$ nodes were drawn from a power-law degree distribution $P(k) \sim k^{-\gamma}$ with cut-off $k_\mathrm{max} < N^{1/2}$. (a) Scaled probability that a degree $k$ node is infected $\rho_k^*/\rho_{k_\mathrm{min}}^*$. (b) Scaled probability of reaching an infected node from a degree $k$ susceptible node $\theta_k^*/\theta_{k_\mathrm{min}}^*$.}
 \label{fig:neighborhood}}
\end{figure}

\modified{
To illustrate how this dichotomy is transposed to degree-based approaches, we present in Fig.~\ref{fig:neighborhood} the behavior of $\rho_k^*$ and $\theta_k^*$ near the absorbing phase $(\lambda \to \lambda_c)$ for quasi-static networks with power-law degree distributions. For an exponent $\gamma = 2.25$, associated with a collective activation, we see in Fig.~\ref{fig:neighborhood}(b) that $\theta_k^*$ is independent of the degree,
\modljd{and $\rho_k^*$ grows linearly with the degree [Fig.~\ref{fig:neighborhood}(a)]. For $\gamma = 3.1$ however, associated with a hub activation, $\theta_k^*$ increases with the degree [Fig.~\ref{fig:neighborhood}(b)], and $\rho_k^*$ grows supra-linearly [Fig.~\ref{fig:neighborhood}(a)]. Our solution [Eq.~\eqref{thetaSol}] reproduces} the qualitative behavior for both scenarios. This indicates that the dichotomy can also be identified and characterized by a degree-based point of view by studying the behavior of $\theta_k^*$ near the absorbing phase. This is achieved with our approach in the following sections.}

\subsection{Perturbative development \label{subsec:perturbative}}

As seen in Fig.~\ref{fig:neighborhood}, the solution for $\theta_k^*$ can be heterogeneous near the absorbing phase. To provide further insights, we consider the absorbing-state limit : we start with an active phase ($\lambda > \lambda_c$), then we take the limit $\lambda \to \lambda_c$, which leads to $\rho_k^*, \theta_k^* \to 0 \fa k$. According to Eq.~\eqref{thetaSol}, to force $\theta_k^* \to 0 \; \forall \; k$, we must require that
\begin{align}\label{conditionAbsorbingState}
	\lim_{\lambda \to \lambda_c} \beta = 0 \quad \text{and} \quad \lim_{\lambda \to \lambda_c} \kappa \geq k_\mathrm{max} \;.
\end{align}
These strong constraints allow us to introduce a perturbative development~: any quantity around the critical threshold is expressed as a power series of $\beta$.

Since the RNA is self-consistent, all quantities [Eqs.~\eqref{defOmega2}, \eqref{thetaSol}, \eqref{parameters}] are interrelated. Therefore, we need to develop them recursively in a coherent way. First, we develop the stationary probability $\theta_k^*$ near the absorbing phase.
\begin{align}
	\theta_k^*(\omega, \lambda) &= \frac{k- \kappa + |k- \kappa| + \frac{2 \alpha \beta (k-1)}{|k - \kappa|} }{2 \alpha (k-1)} + \mathcal{O}(\beta^2) \; \nonumber \\
			   &= \frac{\beta}{\kappa - k} + \mathcal{O}(\beta^2) \;, \label{thetaDev}
\end{align}
where the second equality comes from Eq.~\eqref{conditionAbsorbingState}. However, $\kappa$ also depends on $\beta$ through the quantities ${\Omega^S}^*, {\Omega^S}^*$ and $\Theta^*$. Using Eq.~\eqref{thetaDev} with Eqs.~\eqref{defBigTheta} and \eqref{defOmega2}, we obtain the following leading behaviors
\begin{align*}
	{\Omega^S}^* &= \mathcal{O}(\beta)\;, & {\Omega^I}^* &= \lambda + \mathcal{O}(\beta)\;, & \Theta^* &= \mathcal{O}(\beta)\;.
\end{align*}
This fixes $\kappa$ to order zero, i.e., from Eq.~\eqref{parameters}, we obtain
\begin{align}\label{kappaDev}
	\kappa = \widetilde{\kappa}(\omega,\lambda) + \mathcal{O}(\beta) \;,
\end{align}
where 
\begin{align}\label{kappaDef}
	\widetilde{\kappa}(\omega,\lambda) \equiv \frac{1 + (\lambda + 1)^2 + \omega (2 \lambda + 3) + \omega^2}{\lambda^2} \;.
\end{align}

Combining Eq.~\eqref{kappaDev} with Eq.~\eqref{thetaDev}, we have a coherent development for $\theta_k^*$ 
\begin{align}\label{thetaDev2}
	\theta_k^*(\omega,\lambda) &= \beta f_k(\omega,\lambda) + \mathcal{O}(\beta^2)\;,  
\end{align}
with the auxiliary function
\begin{align}\label{fkDefinition}
	f_k(\omega,\lambda) &\equiv \frac{1}{\widetilde{\kappa}(\omega,\lambda) - k} \;.
\end{align}
Using these definitions, it is possible to express all quantities to first order 
\begin{subequations}\label{MFfirstOrder}
\begin{align}
	{\Omega^S}^* &= \frac{\lambda \avg{f_k k(k-1)}}{\avg{k}} \beta + \mathcal{O}(\beta^2) \;, \\
	{\Omega^I}^* &= \lambda +  \frac{\lambda \avg{f_k^2 k(k-1)}}{\avg{f_k k}} \beta + \mathcal{O}(\beta^2) \;, \\
	\Theta^* &= \frac{\lambda \avg{f_k k^2}}{\avg{k}} \beta + \mathcal{O}(\beta^2) \;.
\end{align}
\end{subequations}
One could continue this perturbative scheme in order to extract the quadratic terms in $\beta$ and so forth. However, the first order development is quite sufficient to characterize the absorbing-state threshold in Sec.~\ref{sec:threshold}.

\subsubsection*{Approximate exponential form}

We can rewrite the solution for $\theta_k^*$ in Eq.~\eqref{thetaDev2} as
\begin{align}\label{exponentialFormDev}
	\theta_k^* &= \frac{\beta}{\widetilde{\kappa}(\omega,\lambda)} \exp \br{-\ln\pr{1- \frac{k}{\widetilde{\kappa}(\omega,\lambda)} }} + \mathcal{O}(\beta^2) \;, \nonumber\\
			   &\approx \frac{\beta}{\widetilde{\kappa}(\omega,\lambda)} \exp \br{\frac{k}{\widetilde{\kappa}(\omega,\lambda)} } \;,
\end{align}
where the approximate exponential form is valid provided $k$ is sufficiently small compared to $\tilde{\kappa}(\omega,\lambda)$. Near the threshold, the density of infected nodes for each degree class is to good approximation $\rho_k^* \approx \lambda k \theta_k^*$ [Eq.~\eqref{rhoSS}]. In the quasi-static limit ($\omega \to 0$) and considering $\lambda \ll 1$, $\tilde{\kappa}(\omega, \lambda \ll 1) \approx 2/\lambda^2$ [Eq.~\eqref{kappaDef}], which leads to the exponential form
\begin{align}\label{exponentialForm}
	\rho_k^* \sim k \exp\pr{\lambda^2 k/2} \;,
\end{align}
This form has been obtained previously by other means in Ref.~\cite{Wei2017}, based upon the results of Ref.~\cite{Boguna2013}. However, they needed to extract $\widetilde{\kappa} \sim \lambda^{-2}$ from numerical simulations, whereas it emerges naturally in our framework. A similar expression has also been found in Ref.~\cite{Ferreira2016} to describe the hub lifetime.

However, the approximate expression Eq.~\eqref{exponentialFormDev} will be inadequate to describe the activity of high degree nodes if $k \sim \tilde{\kappa}(\omega, \lambda)$. In fact, in Sec.~\ref{subsec:threshold_correspondence} we show that the ratio $k_\mathrm{max}/\widetilde{\kappa} \to 1$ near the threshold for a hub dominated phase transition and the development of Eq.~\eqref{exponentialFormDev} breaks down.  


\section{Threshold \label{sec:threshold}}

We now turn our attention towards the absorbing-state threshold $\lambda_c$. Using the perturbative development of Sec.~\ref{subsec:perturbative}, we obtain an explicit upper bound and an implicit expression for $\lambda_c$, which we analytically and numerically compare with existing expressions gathered in Table \ref{tab:threshold}.

\begin{table}
\caption{Threshold estimates for certain formalisms.\label{tab:threshold}}
\begin{ruledtabular}
\begin{tabular}{l l}
Formalism & Threshold estimate $\lambda_c$\\
\hline 
HMF \cite{Boguna2002}	  & $\avg{k}/\avg{k^2}$ \\
PHMF \cite{Mata2014}     & $\avg{k}/\pr{\avg{k^2} - \avg{k}}$ \\
QMF \cite{VanMieghem2009}	  & $1/\mathrm{max}\pr{\sqrt{k_\mathrm{max}}, \avg{k^2}/\avg{k}}$
\end{tabular}
\end{ruledtabular}
\end{table}

\subsection{Explicit upper bound}

An important parameter from the perturbative development is $\widetilde{\kappa}(\omega,\lambda)$, that we call hereafter the \textit{self-activating degree}. In fact, it will become clear throughout the following sections that $\widetilde{\kappa}$ is a good proxy of the minimal degree class able to sustain by itself the dynamics in its neighborhood with correlated reinfections. 

In the absorbing-state limit, Eq.~\eqref{conditionAbsorbingState} leads to the constraint $\widetilde{\kappa}(\omega,\lambda_c) \geq k_\mathrm{max}$. This can be interpreted as follows : the self-activating degree must be higher than the maximal degree, otherwise the system would be in an active phase, sustained by the maximal degree class. This constraint is rewritten as
\begin{align}\label{upperBound}
	\lambda_c(\omega) \leq \frac{1 + \omega + \sqrt{2 k_\mathrm{max} -1 + \omega(3 k_\mathrm{max} -1) + \omega^2 k_\mathrm{max}}}{k_{\mathrm{max}}-1} \;.
\end{align}
Equation \eqref{upperBound} sets a general upper bound on the threshold $\lambda_c$ for any rewiring regime specified by $\omega$. Notably, our approach predicts a vanishing threshold for any random networks with finite $\omega$ in the limit $k_\mathrm{max}\to\infty$.

In the quasi-static limit, we have
\begin{align}\label{upperBoundQS}
	\lambda_c(\omega \to 0) \equiv \lambda_c^{\mathrm{qs}} \leq \frac{1 + \sqrt{2k_{\mathrm{max}} -1 }}{k_{\mathrm{max}}-1} \;.
\end{align}
For large $k_\mathrm{max}$, Eq.~\eqref{upperBoundQS} is well approximated by \mbox{$\lambda_c^{\mathrm{qs}} \lesssim \sqrt{2/k_\mathrm{max}}$}. This upper bound is qualitatively in agreement with QMF (see Table \ref{tab:threshold}) and numerical simulations on static networks \cite{Ferreira2012}. 
\modified{Moreover, Eq.~\eqref{upperBoundQS} can be associated with the threshold of a star graph with $k_\mathrm{max}$ leaves \cite{Cator2013,Mata2013}. This is a natural constraint, since this star is certainly a subgraph of the network due to the presence of $k_\mathrm{max}$ degree nodes. While Eq.~\eqref{upperBoundQS} is slightly different from the threshold suggested by the exact analysis of the star graph \cite{Cator2013}, it is identical to the threshold obtained from PQMF \cite{Mata2013}.} 

In the annealed limit, one expects a finite threshold in the limit $k_\mathrm{max}\to\infty$ for bounded second moment $\avg{k^2}$ \cite{Pastor-Satorras2001}, \modified{i.e for any degree distribution that asymptotically decreases faster than $P(k) \simeq k^{-3}$, in agreement with HMF}. For this condition to be satisfied, Eq.~\eqref{upperBound} prescribes that the rewiring rate $\omega \gtrsim \sqrt{k_\mathrm{max}}$. Therefore, a network with higher degree nodes requires a faster rewiring dynamics to be considered annealed.

\subsection{Self-consistent expression}

Using the definition of $\beta$ in Eq.~\eqref{beta_def} with the first order developments of Eqs~\eqref{MFfirstOrder}, we write the self-consistent expression 
\begin{align}
	\beta = \beta \br{\frac{\pr{\avg{ k(k-1)f_k} + \omega \avg{ k^2 f_k}}(2 + \omega + \lambda)}{\lambda \avg{k}}} + \mathcal{O}(\beta^2) \;,
\end{align}
which can be rewritten as
\begin{align}
	\mathcal{O}(\beta) &=  \pr{\lambda - \frac{(2+\omega)\avg{k f_k}}{(2+\omega)\avg{k^2 f_k} - 2 \avg{k f_k}}} \;.
\end{align}
In the absorbing-state limit, which implies $\beta \to 0$, the term in parentheses on the right must be zero. This defines an implicit expression for the threshold
\begin{align}\label{aaThreshold}
	\lambda_c(\omega) &= \frac{(2+\omega)\avg{k f_k(\omega,\lambda_c)}}{(2+\omega)\avg{k^2 f_k(\omega,\lambda_c)}-2\avg{k f_k(\omega,\lambda_c)}} \;.
\end{align}
Equation \eqref{aaThreshold} is a central result of the RNA---it allows the accurate evaluation of $\lambda_c$ for any degree distribution $P(k)$, and any time scale fixed by $\omega$. For arbitrary $\omega$ and $P(k)$, Eq.~\eqref{aaThreshold} is transcendental and must be solved numerically.

\subsection{Correspondence with existing approaches \label{subsec:threshold_correspondence}}
The transcendental expression for the threshold admits some simplifications for certain limiting cases, leading to many correspondences with current formalisms. First, we consider the extreme regimes of the rewiring process. Equation \eqref{aaThreshold} becomes
\begin{align}\label{aaThresholdLimit}
	\lambda_c &= 
	\begin{dcases}
		\avg{k}/\avg{k^2} & \text{if } \omega \to \infty \;, \\
		\avg{kf_k^{\mathrm{qs}}}/\pr{\avg{k^2f_k^{\mathrm{qs}}}-\avg{kf_k^{\mathrm{qs}}}} & \text{if } \omega \to 0 \;.
	\end{dcases}
\end{align}
where $f_k (\omega \to 0, \lambda_c) \equiv f_k^{\mathrm{qs}} $. Hence, we recover as expected the HMF threshold \cite{Boguna2002} in the annealed limit. In the quasi-static limit, we obtain a threshold similar in form to the one predicted by PHMF, except for the presence of $f_k^{\mathrm{qs}}$ in each average (see Table \ref{tab:threshold}).

To make further progress in the quasi-static limit, let us consider the limit $k_\mathrm{max} \to \infty$. To simplify the notation, we let \mbox{$\widetilde{\kappa}_0 \equiv \widetilde{\kappa}(\omega \to 0, \lambda_c)$}. In this case, there are two possible scenarios for the threshold, depending on the scaling of $\widetilde{\kappa}_0$ with $k_\mathrm{max}$. On the one hand, if \mbox{$\widetilde{\kappa}_0/k_\mathrm{max} \to \infty$}, then $f_k \to \beta/\widetilde{\kappa}_0$, which is independent of the degree. On the other hand, if $\widetilde{\kappa}_0/k_\mathrm{max} \to c \geq 1$, \modified{then $f_k$ depends strongly on the degree} and the threshold $\lambda_c$ is obtained directly. Together, this leads to
\begin{align}\label{aaThresholdLimit2}
	\lambda_c^\mathrm{qs} &= 
	\begin{dcases}
		\avg{k}/\pr{\avg{k^2}-\avg{k}} & \text{if } \widetilde{\kappa}_0/k_\mathrm{max} \to \infty \;, \\
		\sqrt{2}/\sqrt{c k_\mathrm{max}} & \text{if }  \widetilde{\kappa}_0/k_\mathrm{max} \to c \;.
	\end{dcases}
\end{align}

In accordance with the literature \modified{and our previous discussion in Sec.~\ref{subsec:collective_hub}}, we identify the first case in Eq.~\eqref{aaThresholdLimit2} (incidentally the exact same form as the PHMF threshold) with the collective activation scenario. Indeed, since the self-activating degree $\widetilde{\kappa}_0$ is much larger than the maximal degree $k_\mathrm{max}$ just beyond the threshold, none of the degree classes are able to self-sustain the dynamics. The critical phenomenon is therefore truly a collective one. We associate the second case in Eq.~\eqref{aaThresholdLimit2} with the hub activation scenario. Effectively, $\widetilde{\kappa}_0 \sim k_\mathrm{max}$, such that the active phase just beyond the threshold is attributed to the self-activation of the maximal degree class in the network. 
\modified{We can again relate the scaling with $k_\mathrm{max}$ (the second case of Eq.~\eqref{aaThresholdLimit2}) with the threshold of the star graph \cite{Cator2013,Mata2013}. The subgraph containing the hubs and their neighbors (maximal degree stars) is therefore the dominant topological structure responsible for the onset of the active phase.}

\begin{figure}
\includegraphics[width = 0.40\textwidth]{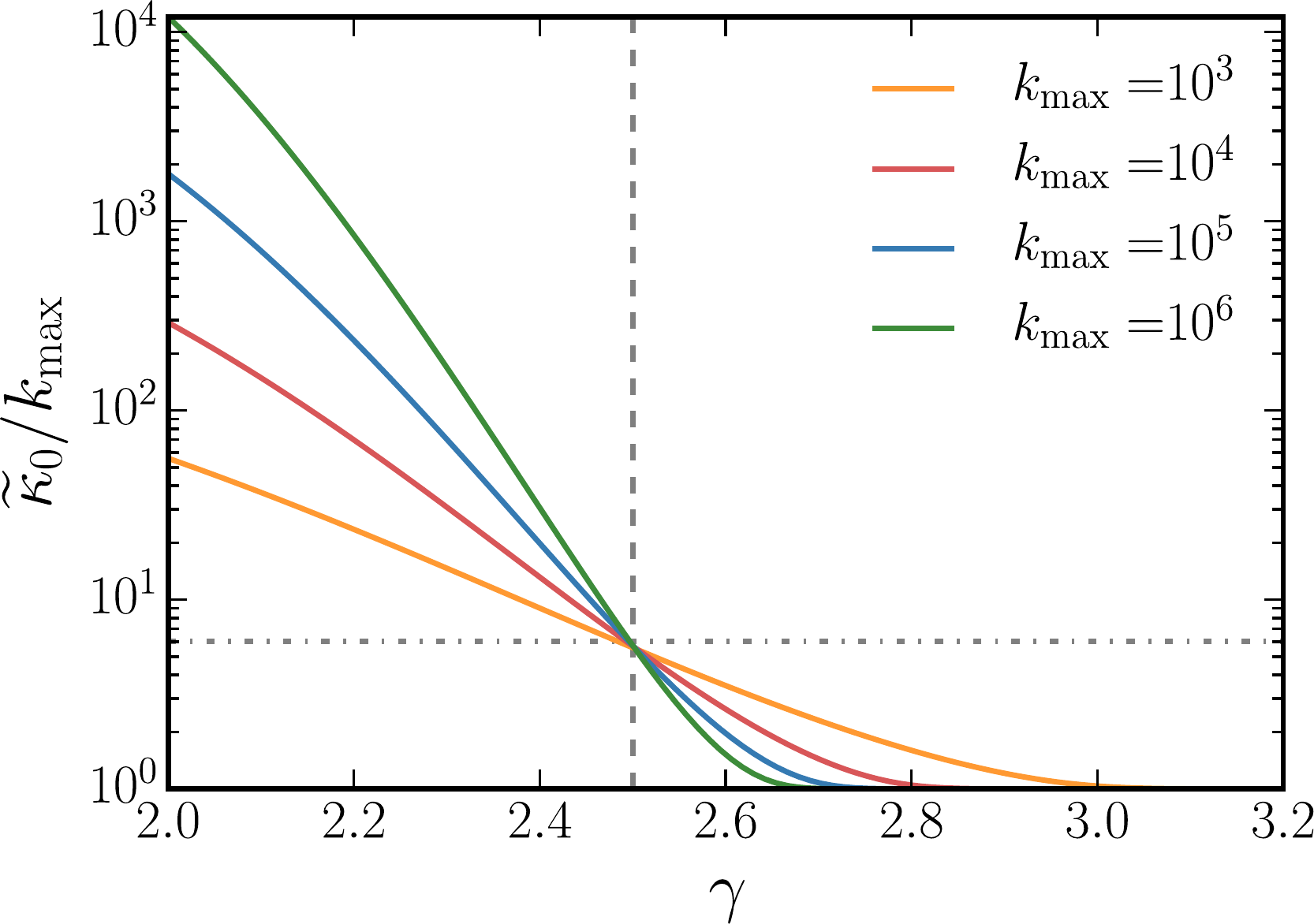}
\caption{(Color online). Ratio $\widetilde{\kappa}_0/k_\mathrm{max}$ against the power-law degree distribution exponent $\gamma$ in the quasi-static limit ($\omega \to 0$), for different values of maximum degree $k_\mathrm{max}$. The minimum degree is \mbox{$k_\mathrm{min} = 3$}. Vertical dashed line corresponds to $\gamma = 5/2$. Horizontal dashed line corresponds to \mbox{$\widetilde{\kappa}_0/k_\mathrm{max} = 2 k_\mathrm{min}$}, identified using Eq.~\eqref{aaThresholdLimit2}. \label{fig:kappa_vs_gamma}}
\end{figure}

This correspondence can be verified explicitly for power-law degree distributions $P(k) \sim k^{-\gamma}$, for which a transition between the collective and hub dominated scenario appears at $\gamma = 5/2$ \cite{Castellano2012,Goltsev2012}. This is done in Fig.~\ref{fig:kappa_vs_gamma} where, as expected, the ratio $\widetilde{\kappa}_0/k_\mathrm{max}$ is a growing function of $k_\mathrm{max}$ for $\gamma < 5/2$, while it goes to 1 for $\gamma > 5/2$---the threshold then coalesces with the upper bound \eqref{upperBoundQS}. This type of result has been observed numerically \cite{Ferreira2012,Mata2013} and is coherent with individual-based approaches \cite{Pastor-Satorras2015}. Precisely at $\gamma = 5/2$, the ratio of the first two moments, $\avg{k^2}/\avg{k}$, is equal to $\sqrt{k_\mathrm{max} k_\mathrm{min}}$, which lead all curves of $\widetilde{\kappa}_0/k_\mathrm{max}$ to cross at the same point $c = 2 k_\mathrm{min}$.

\modified{The two different expressions in Eq.~\eqref{aaThresholdLimit2} are similar to the ones for QMF (see Table.~\ref{tab:threshold}). One is reminded that the QMF estimate for the epidemic threshold is formally a lower bound for the real threshold \cite{VanMieghem2013}, but it is nonetheless qualitatively correct \cite{Ferreira2012}. Therefore, Eq.~\eqref{aaThreshold} has the appropriate behavior in both the annealed and quasi-static limits. This is further validated with numerical simulations (see Figs.~\ref{fig:PL_threshold} and \ref{fig:RRN_threshold}).}

\subsection{Comparison with simulations}

\begin{figure}
\includegraphics[width = 0.5\textwidth]{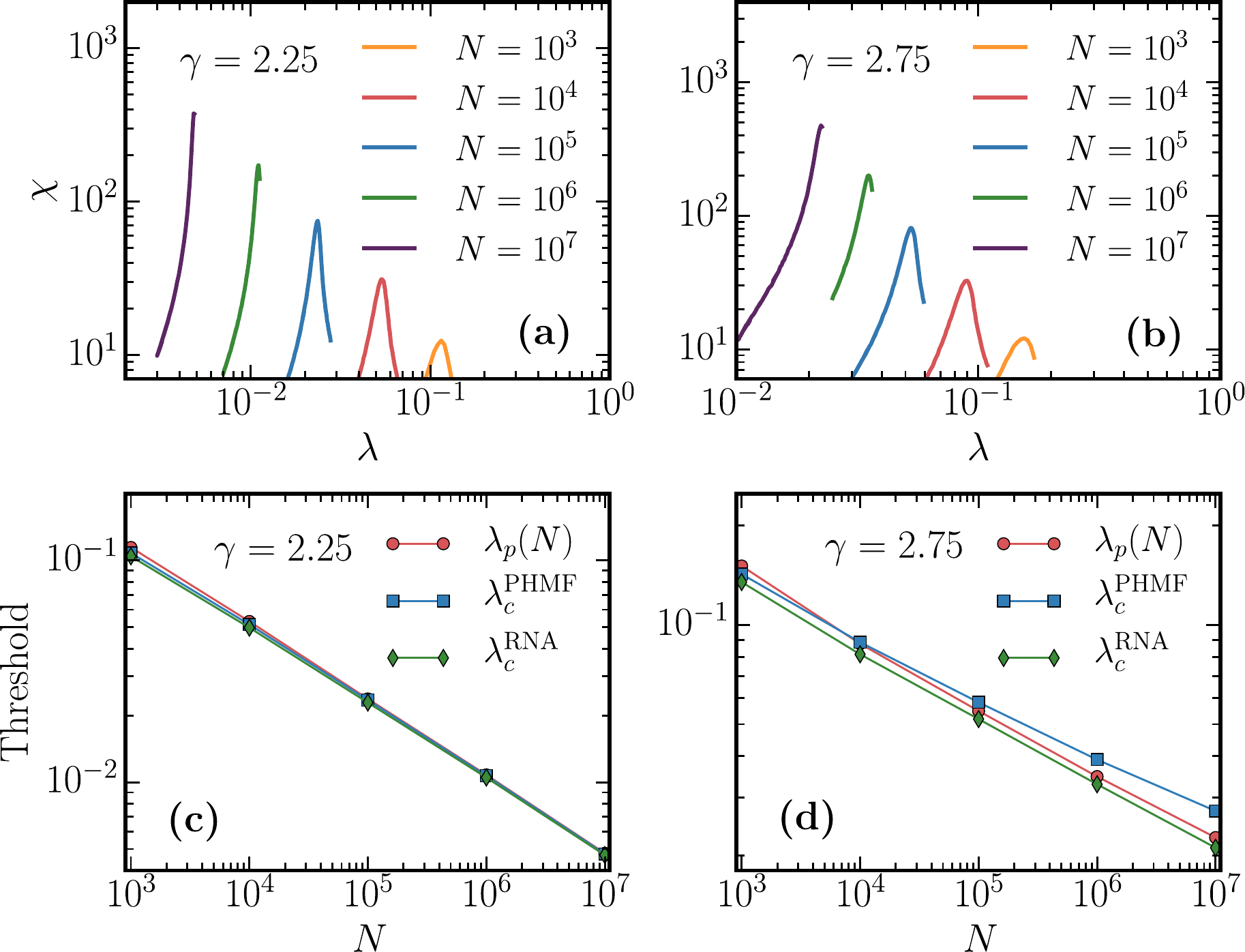}%
\caption{ (Color online). Threshold evaluation for power-law random networks of degree distribution $P(k) \sim k^{-\gamma}$, minimum degree $k_\mathrm{min} = 3$ and maximum degree $k_\mathrm{max} \leq N^{1/2}$. (a)--(b) Susceptibility against the infection rate for a single network realization. (c)--(d) Threshold against the number of nodes (averaged over 10 network realizations) estimated by~: the position of the susceptibility peak $\lambda_p(N)$, our threshold estimate $\lambda_c^\mathrm{RNA}$ of Eq.~\eqref{aaThresholdLimit} for $\omega \to 0$ and the PHMF threshold $\lambda_c^\mathrm{PHMF}$. \label{fig:PL_threshold}}
\end{figure}

We expect that Eq.~\eqref{aaThreshold} should be a good approximation of $\lambda_c$ for finite size realizations of the configuration model with large $N$. This can be verified by sampling the configurations of the system that do not fall on the absorbing state, the \emph{quasi-stationary distribution} \cite{Oliveira2005,Ferreira2011,Sander2016}, to evaluate the susceptibility
\begin{align}
\chi &= \frac{E[n^2]-E[n]^2}{E[n]} \;,
\end{align}
with $n \leq N$ the number of infected nodes in the system and $E[\cdots]$ denotes the expectation over the quasi-stationary distribution. The susceptibility exhibits a sharp maximum at $\lambda_p(N)$ as shown in Fig.~\ref{fig:PL_threshold}(a) and \ref{fig:PL_threshold}(b), corresponding to the epidemic threshold of the system in the thermodynamic limit \cite{Ferreira2012}. 

We have first validated Eq.~\eqref{aaThreshold} regarding the two possible activation schemes using a power-law degree distribution $P(k)\sim k^{-\gamma}$ in the quasi-static limit. Figures \ref{fig:PL_threshold}(c) and \ref{fig:PL_threshold}(d) show that the RNA yields a threshold in agreement with the susceptibility for both the collective ($\gamma \leq 5/2$) and the hub dominated ($\gamma > 5/2$) phase transition. As a comparison, it is seen in Fig.~\ref{fig:PL_threshold}(d) that the prediction of PHMF does not reproduce the scaling of $\lambda_p(N)$ for the hub activation scenario. This is explained by the fact that this approach neglects the neighborhood heterogeneity. Despite being accurate for collective activation \cite{Mata2014}, as seen in Fig.~\ref{fig:PL_threshold}(c), PHMF is unable to describe correctly a hub dominated dynamics.

\begin{figure}
\centering
\includegraphics[width = 0.40\textwidth]{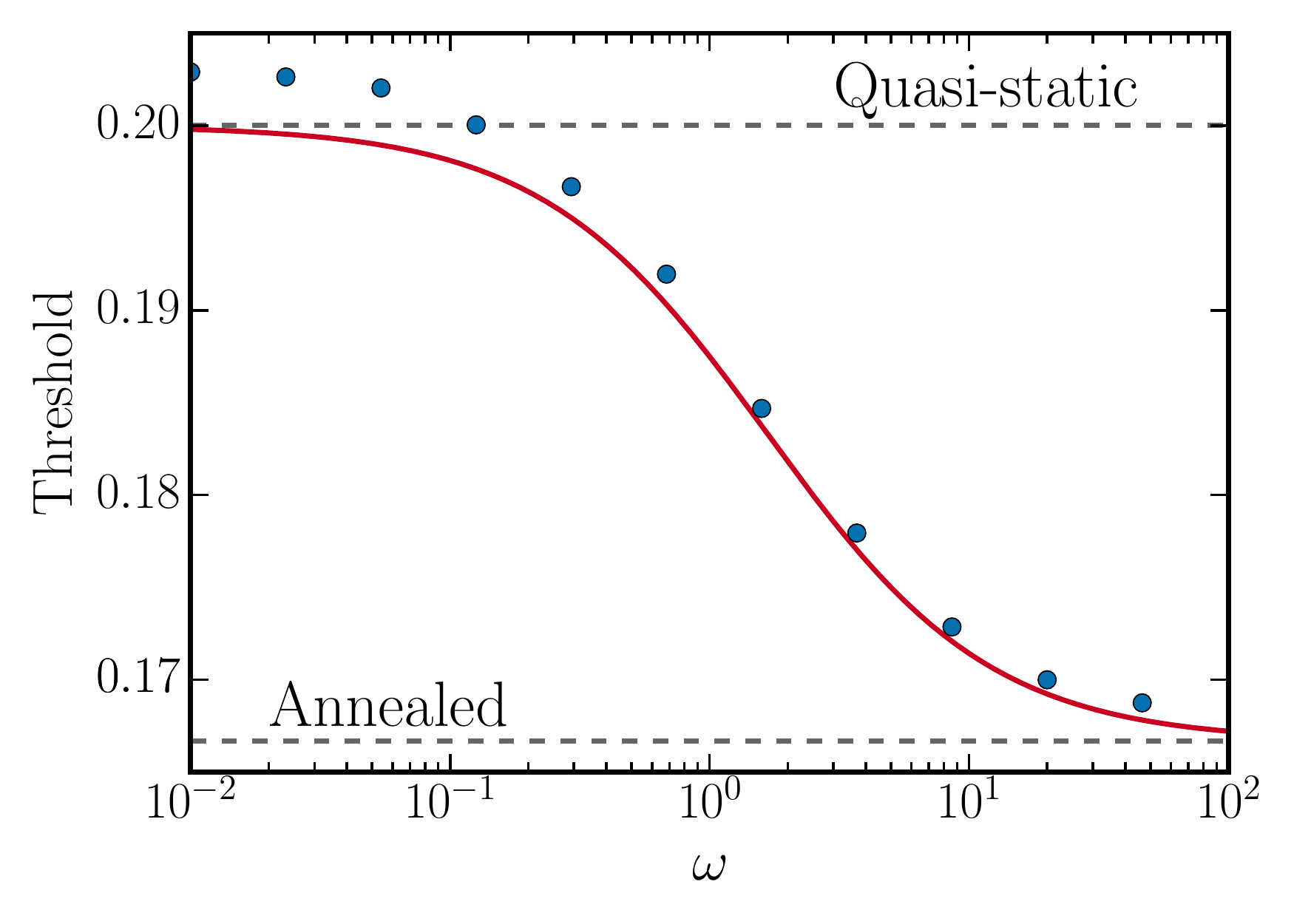}%
\caption{(Color online). Threshold against the rewiring rate for a regular random network with degree $k_0 = 6$ and network size $N = 10^5$. The solid line represents the threshold estimated by Eq.~\eqref{threshold_RRN} and the markers represent the positions of the susceptibility peaks $\lambda_p(N)$. The disparity with the simulations is attributed to a combination of finite size effects and approximations leading to Eq.~\eqref{aaThreshold}. \label{fig:RRN_threshold}}
\end{figure}

Moreover, Eq.~\eqref{aaThreshold} is versatile and predicts the threshold for all intermediate regimes between the annealed and quasi-static limit. To illustrate this feature, we have extended the standard quasi-stationary distribution method to include the rewiring procedure (see Appendix \ref{app:monte_carlo}). For the sake of simplicity, we have applied it to a regular random network with distribution $P(k) = \delta_{kk_0}$, for which Eq.~\eqref{aaThreshold} yields the threshold
\begin{align}\label{threshold_RRN}
	\lambda_c(\omega) &= \frac{2 + \omega}{(2+\omega)k_0 - 2} \;.
\end{align}
The validation is presented in Fig.~\ref{fig:RRN_threshold}. Equation \eqref{threshold_RRN} reproduces with good accuracy the smooth transition from one regime to another.

\subsection{Non-monotonicity of the threshold}

Equation \eqref{threshold_RRN} and Fig.~\ref{fig:RRN_threshold} suggest a monotically decreasing threshold with growing rewiring rate $\omega$. One may ask: is this always the case? Equation \eqref{aaThreshold} is much more intricate and does not possess an explicit dependence upon $\omega$ for general degree distributions. 

To answer this question, it is important to note that the random rewiring of the edges affects the threshold in two different ways. On the one hand, it promotes the contact between infected and susceptible nodes (the dynamic correlation is reduced), which decreases the threshold (see Fig.~\ref{fig:RRN_threshold}). On the other hand, random rewiring inhibits the reinfection of hubs by their neighbors, which is driving the hub dominated phase transition.

\begin{figure}
\includegraphics[width = 0.40\textwidth]{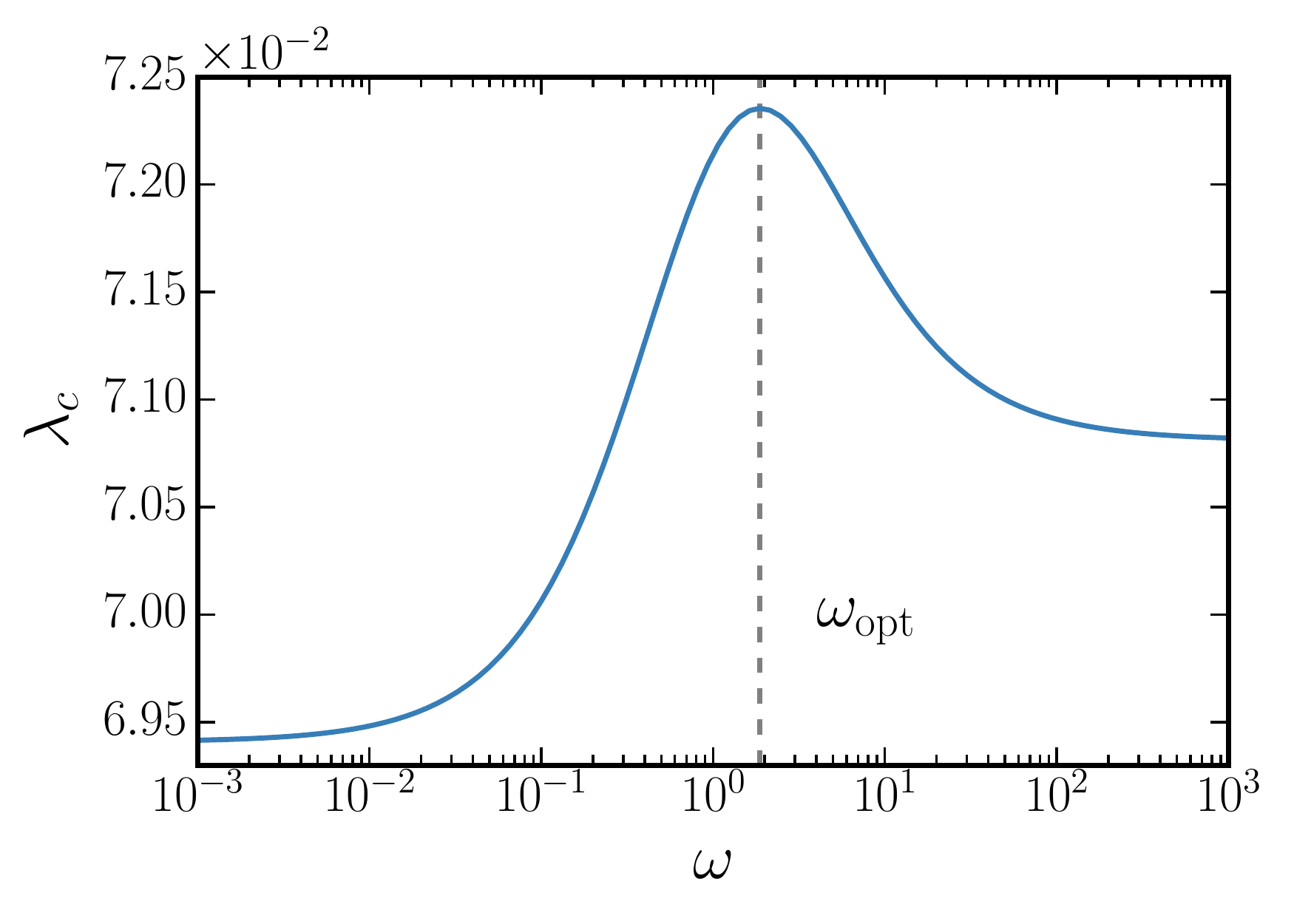}%
\caption{Threshold against the rewiring rate for power-law degree distribution $P(k) \sim k^{-\gamma}$ with exponent $\gamma = 2.75$ and maximal degree $k_\mathrm{max} = 50$. The threshold is evaluated with Eq.~\eqref{aaThreshold}. The dashed line marks the optimal value $\omega_\mathrm{opt}$ at which $\lambda_c$ is maximized. \label{fig:optimal_rewiringRate}}
\end{figure}

For heterogeneous networks that are affected by both mechanisms, this leads to a non-monotonic relation for $\lambda_c(\omega)$, as presented in Fig.~\eqref{fig:optimal_rewiringRate}. There exists a value $\omega_\mathrm{opt}$ at which $\lambda_c(\omega)$ is maximized : the hub reinfection mechanism is inhibited, without too much stimulating the spreading through new infected-susceptible contacts. The value $\omega_\mathrm{opt}$ then defines the optimal rewiring rate to hinder the infection spreading on a network with a specified degree distribution.

\section{Critical exponents \label{sec:critical_exponent}}

To complete the phase transition portrait, we address the theoretical determination of the critical exponents of $\rho^*$, the mean infected density, and $\theta_k^*$, which describes the neighborhood for each degree class. More specifically, we characterize the scaling exponents $\delta$ associated with
\begin{align}
	\rho^* \sim (\lambda - \lambda_c)^\delta \;,
\end{align}
and $\cbrace{\eta_k}$ related to
\begin{align}
	\theta_k^* \sim (\lambda -\lambda_c)^{\eta_k} \;.
\end{align}

To make analytical progress, we restrict ourselves to power-law degree distribution $P(k) = A k^{-\gamma}$ in the limit $k_\mathrm{max} \to \infty$. The case $\omega \to \infty$, the annealed limit, has already been analyzed through the HMF framework \cite{Pastor-Satorras2001} and leads to the following critical exponents
\begin{align}\label{HMF_exponents}
\delta^{\mathrm{HMF}} &= \begin{dcases}
						 1/(3-\gamma) & \text{for } \gamma < 3 \;, \\
						 1/(\gamma - 3) & \text{for } 3 <\gamma < 4 \;, \\
						 1 & \text{for } \gamma \geq 4 \;,
						\end{dcases} \\
\eta_k^{\mathrm{HMF}} &= \begin{dcases}
						 (\gamma -2)/(3-\gamma) & \text{for } \gamma < 3 \;, \\
						 1/(\gamma - 3) & \text{for } 3 <\gamma < 4 \;, \\
						 1 & \text{for } \gamma \geq 4 \;,
						\end{dcases} \;
\end{align}
with $\eta_k$ being the same $\forall \; k$. Note that for $\gamma > 3$, $\lambda_c > 0$ for annealed networks. 

In this section, we consider the case study of finite $\omega$, leading to a vanishing threshold $\lambda_c \to 0$ for all degree distribution exponents $\gamma$ in the limit $k_\mathrm{max} \to \infty$ [see Eq.~\eqref{upperBound}].

\subsection{Bounds on the critical exponents \label{subsec:bound}}

The solution for $\theta_k^*$ in Eq.~\eqref{thetaSol} has a complicated dependence on each degree class and is ill suited for the direct estimation of the critical exponents. Instead, we consider lower and upper bounds for various quantities near the absorbing phase, each identified by the subscript ``${-}$'' or ``${+}$'' respectively. For instance, $\theta_{-}^*$ and $\theta_{+}^*$ are lower and upper bounds for $\theta_k^*$ respectively, valid for all degree classes. 

We are mostly interested in the scaling of these quantities with $\lambda$ near the absorbing phase, hence lower and upper bounds are expressed only up to a constant factor. According to Eq.~\eqref{thetaSol}, we can set the following bounds for $\theta_k^*$ (see Appendix \ref{app:lower_upper_bound} for details)
\begin{subequations}
\begin{align}
	\theta_{{-}}^* &\equiv \br{\frac{\beta}{\kappa}}_{-} \sim {\Omega^S}^*_{-} + \omega \Theta^*_{-} \;, \\
	\theta_{{+}}^* &\equiv \br{\frac{1}{\alpha}}_{+} \sim {\Omega^I}^*_{+} + \omega \Theta^*_{+} \;,
\end{align}
\end{subequations}
\modified{The bracket $[x]_{-/+}$ indicates that we take the lower/upper bound of $x$}. This permits us to obtain bounds for other quantities in terms of the bounds for $\theta_k^*$---for instance ${\Omega^S}_{-}^*$ in terms of $\theta_{-}^*$, leading to self-consistent expressions. 

Since the developments for lower and upper bounds are the same, we write explicit equations in terms of $\theta_\pm^*$. For ${\Omega^S}^*$, according to Eq.~\eqref{defOmega2}, this leads to
\begin{align}\label{integralFormOmegaMin}
	{\Omega^S}_\pm^* =& \frac{\lambda(1-\theta_\pm^*)}{\avg{k}} \left[ A \theta_\pm^* \int_{k'}^\infty \frac{\pr{k^{2- \gamma} - k^{1- \gamma}}}{1 + \lambda \theta_\pm^* k} \dx k\right.  \nonumber \\ 
	&+ \theta_\pm^* \avg{(k-1)k}_{k'} \left. \vphantom{\int_{k'}^\infty} \right] +\mathcal{O}\pr{\lambda^2 {\theta_\pm^*}^2 } \;,
\end{align}
where $\avg{ \cdots}_{k'}$ represents an average over $P(k)$ from $k_\mathrm{min}$ to $k'-1$, and $k'$ is a finite value chosen such that the rest of the average can be approximated by an integral. 

For $\lambda \theta_\pm^* \to 0$, we can then extract the leading terms of the integral in Eq.~\eqref{integralFormOmegaMin} (see Appendix \ref{app:integral_evaluation}). This leads to
\begin{align}\label{asymptoticOmegaS}
 	{\Omega^S}_\pm^* =& (1- \theta_\pm^*) \left[ \vphantom{(\lambda \theta_\pm^*)^{\gamma - 2}} \right. a_1(\lambda \theta_\pm^*)^{\gamma - 2} + a_2 \lambda \theta_\pm^* \nonumber \\
 	&+ a_3 (\lambda \theta_\pm^*)^{\gamma - 1} \left. \vphantom{(\lambda \theta_\pm^*)^{\gamma - 2}} \right]+ \mathcal{O}(\lambda^2 {\theta_\pm^*}^2) 
 \end{align}
Similarly, using Eq.~\eqref{defOmega2} and \eqref{defBigTheta}, we obtain
\begin{align}
	{\Omega^I}_\pm^* =& \lambda + b_1\frac{(\lambda \theta_\pm^*)^{\gamma - 1}}{\rho_\pm^*} + b_2 \frac{\lambda^2 {\theta_\pm^*}^2}{\rho_\pm^*} \nonumber \\ 
	&+ b_3 \frac{(\lambda \theta_\pm^*)^{\gamma}}{\rho_\pm^*} + \mathcal{O}\pr{\frac{\lambda^3 {\theta_\pm^*}^3}{\rho_\pm^*}} \;, \label{asymptoticOmegaI}\\
	\Theta_\pm^* =& c_1 (\lambda \theta_\pm^*)^{\gamma - 2} + c_2 \lambda \theta_\pm^* + \mathcal{O}(\lambda^2 {\theta_\pm^*}^2)\;, \label{asymptoticTheta}\\
	\rho_\pm^* =& d_1 \lambda \theta_\pm^* + d_2 (\lambda \theta_\pm^*)^{\gamma - 1} + \mathcal{O}(\lambda^2 {\theta_\pm^*}^2) \label{asymptoticRho}
\end{align}
where the coefficients $\cbrace{a_i,b_i,c_i,d_i}$ are non-vanishing constants in the absorbing-state limit. We now consider separately the region $2 < \gamma < 3$ and $\gamma \geq 3$.

\subsubsection{Region $2 < \gamma < 3$}

Since ${\Omega^S}^*$ and $\Theta^*$ possess the same critical behavior according to Eqs.~\eqref{asymptoticOmegaS} and \eqref{asymptoticTheta}, the lower bound $\theta^*_{-}$ possesses the simple self-consistent expression
\begin{align}
	\theta^*_{-} \sim (\lambda \theta^*_{-})^{\gamma-2} \implies \theta^*_{-} \sim \lambda^{(\gamma - 2)/(3- \gamma)} \;.
\end{align}
Combining this with Eq.~\eqref{asymptoticRho}, we obtain
\begin{align}\label{lowerBoundRho_regime1}
	\rho^*_{-} \sim \lambda^{1/(3- \gamma)} \equiv \lambda^{\delta_{+}}\;.
\end{align}
The upper bound is slightly more complicated : ${\Omega^I}^*$ and $\Theta^*$ might not possess the same critical behavior. However, by definition we know that ${\Omega^I}^* \geq {\Omega^S}^* \sim \Theta^*$, hence ${\Omega^I}^*$ is always dominant for finite rewiring rates $\omega$. This implies that a finite rewiring rate does not have any impact on the critical exponents. We therefore have
\begin{align}
	\theta^*_{+} \sim {\Omega^I}^*_{+} \simeq \lambda + b_1\frac{(\lambda \theta^*_{+})^{\gamma - 1}}{\rho^*_{+}} \;.
 \end{align}
Using Eq.~\eqref{asymptoticRho}, we obtain
\begin{align}\label{upperBoundRho_regime1}
	\theta_{+}^* &\sim \lambda^\psi\;, & \rho^*_{+} &\sim \lambda^{\psi +1} \equiv \lambda^{\delta_{-}} \;.
\end{align}
where
\begin{align}
\psi = \begin{dcases}
			\frac{\gamma-2}{3- \gamma} & \text{for } \gamma \leq 5/2 \;, \\
			1 & \text{for } \gamma > 5/2 \;.
			\end{dcases}
\end{align}

\begin{figure}[tb]
	\centering
	\includegraphics[width = 0.4\textwidth]{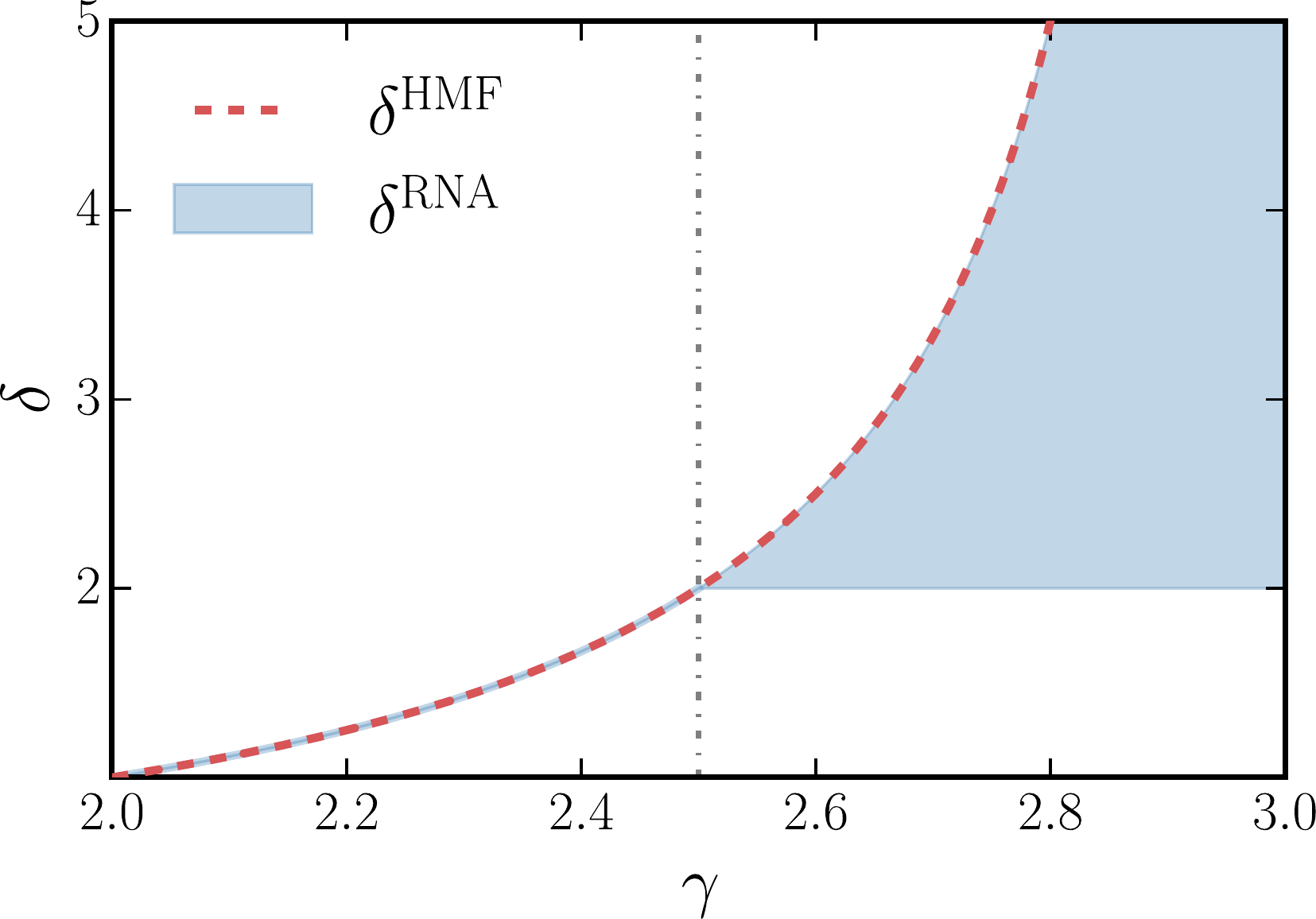}
	\caption{(Color online). Critical exponent $\delta$ associated to the mean infected density $\rho^*$, for a power-law degree distribution $P(k) \sim k^{-\gamma}$ in the thermodynamic limit. The bounded (shaded) region and the solid line correspond to the exponent predicted by our approach \modified{[Eqs.~\eqref{lowerBoundRho_regime1} and \eqref{upperBoundRho_regime1}]} and the dashed line to the HMF exponent. The dashed-dotted line indicates the transition from the collective to the hub dominated region ($\gamma = 5/2$).}
	\label{fig:bounded_region_meanInfectedDensity}
\end{figure}

Equations \eqref{lowerBoundRho_regime1} and \eqref{upperBoundRho_regime1} fix the bounds for the critical exponent $\delta$, as presented in Fig.~\ref{fig:bounded_region_meanInfectedDensity}. In the region $\gamma \leq 5/2$, associated to the collective activation scheme, upper and lower bounds collapse to the annealed exponent of Eq.~\eqref{HMF_exponents}, namely $\delta= 1 / (3-\gamma)$. This is in fact the region where the annealed regime describes the dynamics well, even for static networks \cite{Ferreira2012}. 

However, in the hub activation region ($\gamma > 5/2$), the bounds are different, $\delta_{+}= 1 /(3-\gamma)$, $\delta_{-} = 2$, giving rise to a wide range for the values of the critical exponent. We will see in Sec.~\ref{subsec:HCP} that this behavior is related to the emergence of a \textit{heterogeneous critical phenomenon} in this region. Nevertheless, it is straightforward to verify that these bounds are not in contradiction with the exact ones ($\gamma-1 \leq \delta \leq 2 \gamma - 3$) of Ref.~\cite{Chatterjee2009} for static networks.

\subsubsection{Region $\gamma \geq 3$}

The lower bound $\theta^*_{-}$ in this region can be determined again using $\theta^*_{-} \sim {\Omega^S}^*_{-} + \omega \Theta^*_{-}$. More explicitly, in this region we have
\begin{align}
	\theta^*_{-} \simeq e_1 \lambda \theta^*_{-} + e_2(\lambda \theta^*_{-})^{\gamma - 2} - e_3 \lambda (\theta^*_{-})^2 \;,
\end{align}
where $\cbrace{e_i}$ are non-vanishing constants formed by the combination of $\cbrace{a_i,c_i}$. This leads to a critical behavior of the form
\begin{align}
	\rho^*_{-} \sim \theta^*_{-} \sim \pr{\lambda - \lambda_\mathrm{e}}^\nu \;,
\end{align}
where $\nu = \mathrm{max}\br{1, 1/(\gamma-3)}$. Therefore, the lower bound is associated with a finite effective threshold defined by \mbox{$\lambda_\mathrm{e} \equiv e_3^{-1}$}. This is at odds with the upper bound in this region, which is the continuity of the previous region
\begin{align}
	\theta^*_{+} &\sim \lambda \;, & \rho^*_{+} &\sim \lambda^2\;.
\end{align}
In brief, the two bounds are even more separated from each other in this region.

\subsection{Heterogeneous critical phenomenon \label{subsec:HCP}}

Using the results of Sec.~\ref{subsec:bound}, it is also possible to get some insight on the critical behavior of $\theta_k^*$ for extreme degree classes, $\theta_{k_\mathrm{min}}^*$ and $\theta_{k_\mathrm{max}}^*$ (the limit $k_\mathrm{max} \to \infty$ is still implicitly considered). We stress that $\theta_{k_\mathrm{min}}^*$ and $\theta_{k_\mathrm{max}}^*$ are different from $\theta^*_{-}$ and $\theta^*_{+}$.

\begin{figure}[tb]
	\centering
	\includegraphics[width = 0.4\textwidth]{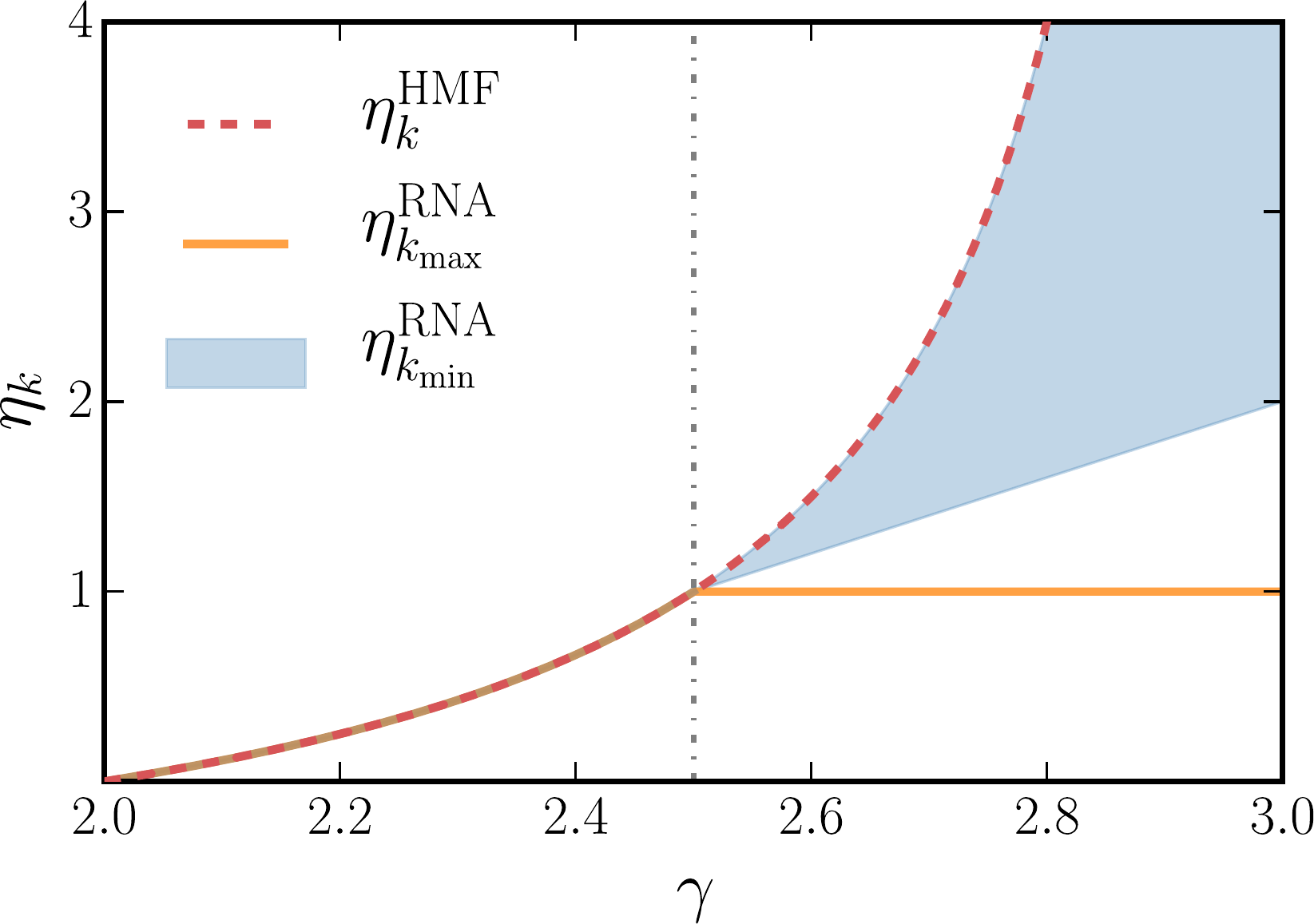}
	\caption{(Color online). Critical exponents $\eta_k$ associated to $\theta_k^*$, for power-law degree distribution $P(k) \sim k^{-\gamma}$ in the thermodynamic limit. The bounded (shaded) region and the solid lines correspond to the exponents predicted by our approach \modified{[Eqs.~\eqref{exponent_eta_kmin} and \eqref{exponent_eta_kmax}]} and the dashed line to the HMF exponent. The dashed-dotted line indicates the transition from the collective to the hub dominated region ($\gamma = 5/2$).}
	\label{fig:HCP}
\end{figure}

\begin{figure}
\includegraphics[width = 0.5\textwidth]{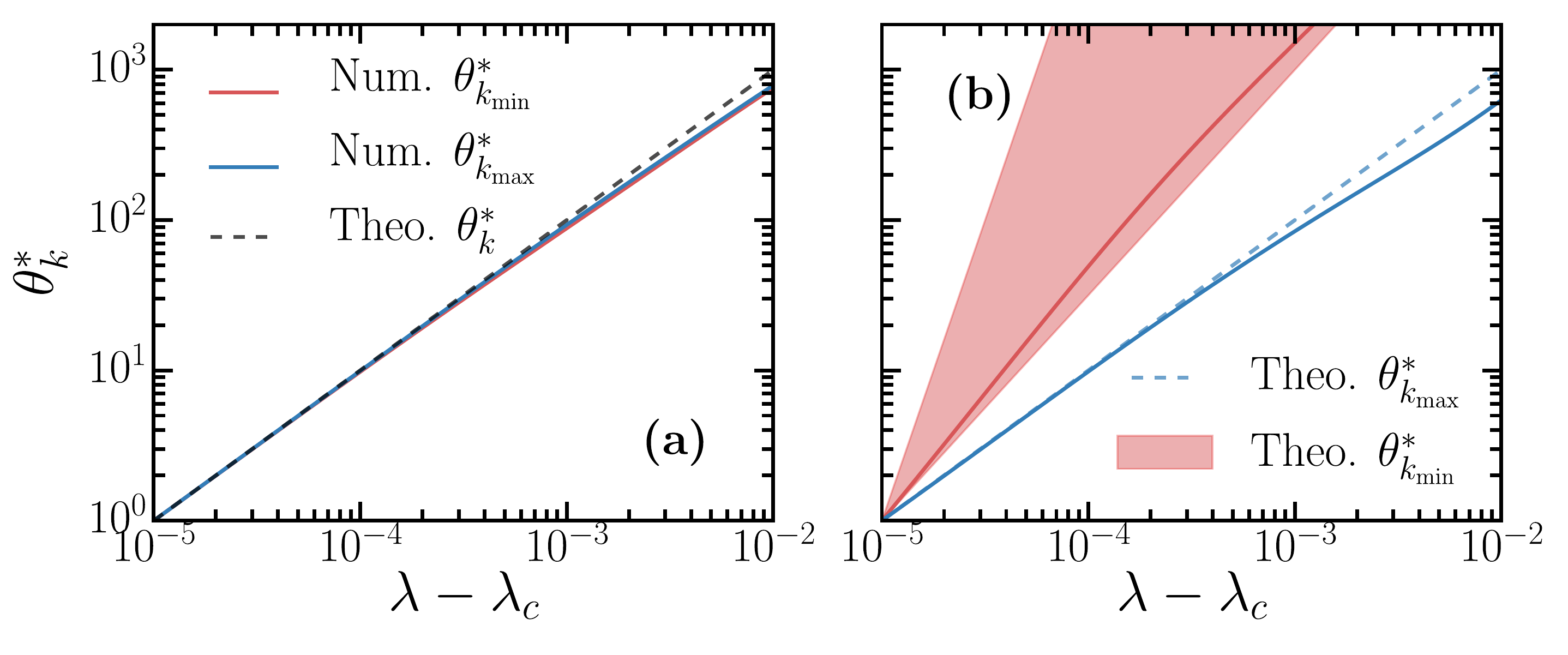}%
\caption{(Color online). Critical behavior for $\theta_{k_\mathrm{min}}^*$ and $\theta_{k_\mathrm{max}}^*$. All curves have been normalized to the value of $\theta_k^*$ at the origin of the abscissae. Dashed lines and shaded regions correspond to the theoretical scaling predicted in the limit $k_\mathrm{max} \to \infty$. Solid lines represent the numerical evaluation of Eq.~\eqref{thetaSol} for bounded degree distribution with $k_\mathrm{min} = 3$ and $k_\mathrm{max} = 5\times 10^5$. (a) Power-law degree distribution with exponent $\gamma = 2.5$. (b) Power-law degree distribution with exponent $\gamma = 2.75$. \label{fig:scaling_thetak}}
\end{figure}

According to Eq.~\eqref{thetaSol}, we have the following behavior near the absorbing phase (see Appendix \ref{app:theta_critical_behavior} for details)
\begin{subequations}
\begin{align}
	\theta_{k_\mathrm{min}}^* &\simeq \frac{\beta}{\kappa} \sim {\Omega^S}^* + \omega \Theta^* \;, \\
	\theta_{k_\mathrm{max}}^* &\simeq \frac{1}{\alpha} \sim {\Omega^I}^* + \omega \Theta^* \;.
\end{align}
\end{subequations}
Using the expressions for $\theta^*_{-}$ and $\theta^*_{+}$ to bound ${\Omega^S}^*$ and ${\Omega^I}^*$, we arrive at the following portrait
\begin{align}
	\theta_{k_\mathrm{min}}^* &\lesssim \lambda^{(\psi + 1)(\gamma - 2)} \;, \\
	\theta_{k_\mathrm{min}}^* &\gtrsim \theta^*_{-} \;, \\
	\theta_{k_\mathrm{max}}^* &\sim \lambda^{\psi} \;,
\end{align}
which characterizes the critical exponents $\eta_{k_\mathrm{min}}$ and $\eta_{k_\mathrm{max}}$. For instance, for $2 < \gamma < 3$, we have
\begin{align}\label{exponent_eta_kmin}
	\mathrm{min}\br{2\gamma - 4, \frac{\gamma-2}{3- \gamma} }\leq \eta_{k_\mathrm{min}} \leq \frac{\gamma-2}{3- \gamma} \;,
\end{align}
and
\begin{align}\label{exponent_eta_kmax}
	\eta_\mathrm{max} = \mathrm{min}\br{1, \frac{\gamma-2}{3- \gamma}} \;.
\end{align}
It is a striking new result : as presented in Fig.~\eqref{fig:HCP}, in the hub dominated regime ($\gamma > 5/2$), the bounded regions for $\eta_{k_\mathrm{min}}$ and $\eta_{k_\mathrm{max}}$ are disjoint. These different asymptotic scalings are validated for finite $k_\mathrm{max}$ in Fig.~\ref{fig:scaling_thetak}.

Different critical exponents for extreme degree classes is also an elegant explanation for the heterogeneity of $\theta_k^*$ observed in Fig.~\ref{fig:neighborhood}(b). Indeed, near the absorbing phase,
\begin{align}\label{thetak_ratio}
	\frac{\theta_{k_\mathrm{min}}^*}{\theta_{k_\mathrm{max}}^*} \sim \lambda^{\eta_{k_\mathrm{min}} - \eta_{k_\mathrm{max}}} \equiv \lambda^{\Delta} \;,
\end{align}
with $\Delta > 0$ for $\gamma > 5/2$. Moreover, it illustrates that the critical phenomenon is itself \textit{heterogeneous}, involving different mechanisms depending on the degree class : for hubs, activity is supported locally through correlated reinfections, while for the rest of the system, activity is mostly due to the propagation induced by the hubs. 

This results also have an impact on how $\rho_k^*$ grows for each degree class beyond $\lambda_c$, according to Eq.~\eqref{rhoSS}. It explains the wide bounds we obtained for $\rho^* = \avg{\rho_k^*}$ in the hub activation region, since $\rho_k^*$ grows differently for each degree class. 


\section{Beyond the hub activation threshold \label{sec:beyond}}

As presented in Sec.~\ref{subsec:threshold_correspondence}, a collective activation leads to $\theta_k^* \sim f_k$ \modified{independent of the degree, while a hub activation results in a growing function of the degree} (see Fig.~\ref{fig:neighborhood}). The latter is formally identified as a heterogeneous critical phenomenon [Eq.~\eqref{thetak_ratio}]. However, this analysis based on the critical exponents is well defined only in the combined limit $k_\mathrm{max} \to \infty$ and \mbox{$\lambda \to 0$}, in which case the impact of the rewiring is lost. 

\begin{figure}
\includegraphics[width = 0.40\textwidth]{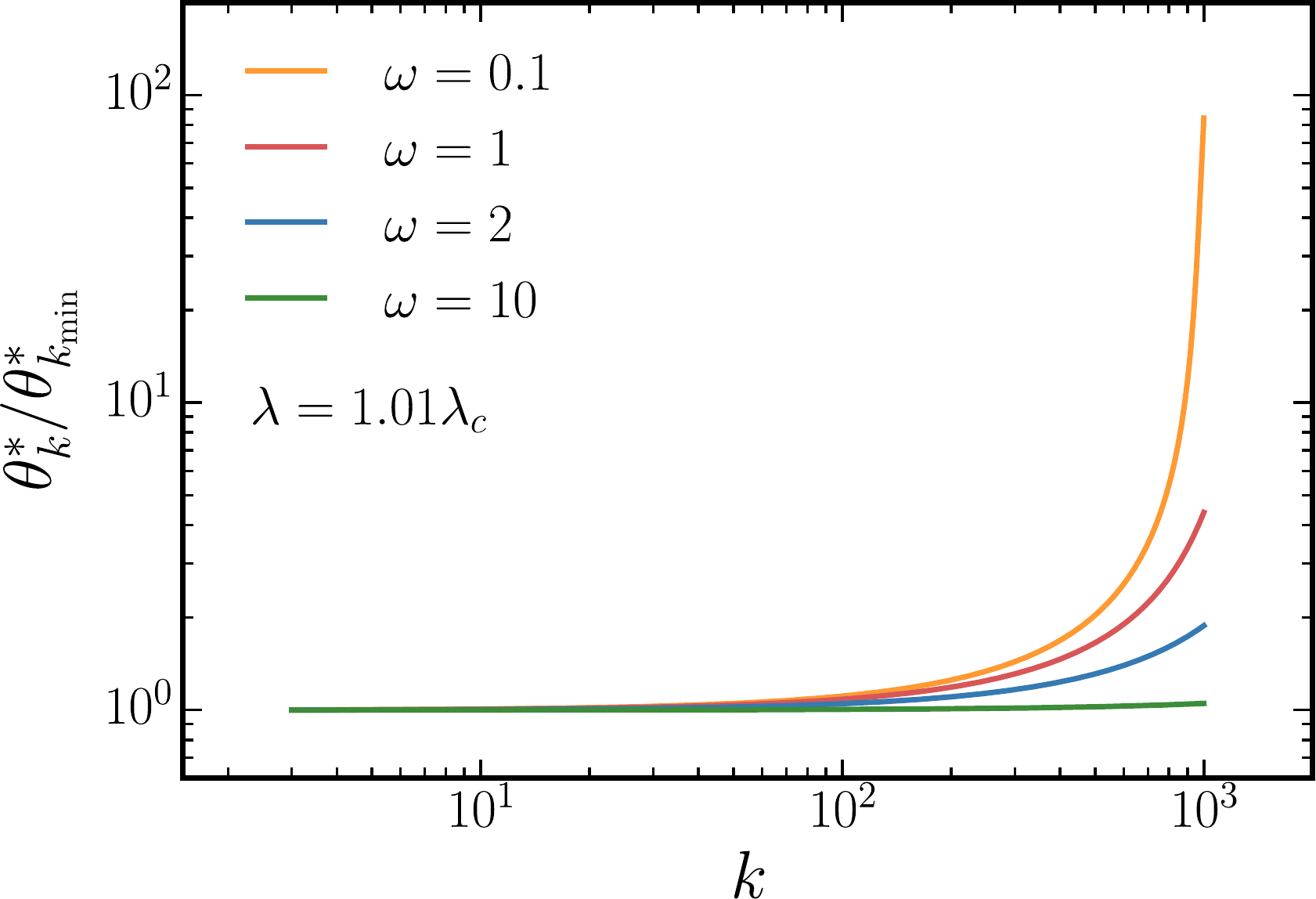}%
\caption{(Color online). $\theta_k^*/\theta_{k_\mathrm{min}}^*$ near the absorbing phase for power-law degree distribution with exponent \mbox{$\gamma = 3.1$} and maximal degree $k_\mathrm{max} = 1000$, for different values of rewiring rates $\omega$. The minimal degree is $k_\mathrm{min} = 3$. \label{fig:thetak_omega}}
\end{figure}

Beyond the threshold and for finite $k_\mathrm{max}$, the dichotomy is not as well defined and the rewiring rate $\omega$ does have a significant impact. In fact, the structural dynamics permits us to interpolate between the two scenarios. According to Eq.~\eqref{kappaDef}, the rewiring rate $\omega$ increases the self-activating degree $\widetilde{\kappa}(\omega,\lambda)$, forcing a more collective activation. This leads to a more homogeneous neighborhood among the degree classes near the absorbing phase, as seen in Fig.~\ref{fig:thetak_omega}. 

Also, critical exponents of Sec.~\ref{sec:critical_exponent} do not inform us on the behavior of the system far beyond the hub activation threshold. For power-law degree distribution having an exponent $\gamma > 3$, it has been observed in numerical simulations that the \textit{delocalization} of the dynamics, where not only hubs sustain the propagation, happens at a finite $\lambda$. This gives rise to a second peak on the susceptibility curve $\chi$, associated with the \modified{activation of the shell with the largest index in the $K$-core decomposition} \cite{Ferreira2012} and seems to correspond with the HMF threshold \cite{Mata2015}.

Our compartmental formalism is not well suited to identify precisely this second transition. However, we are able to describe how the system behaves as the infection rate is increased beyond $\lambda_c$, towards this delocalized regime. An interesting feature is the \textit{successive activation} of the degree classes. According to Eq.~\eqref{kappaDef}, the self-activating degree $\widetilde{\kappa}$ is a monotically decreasing function of $\lambda$. Since $\widetilde{\kappa}(\omega,\lambda_c) \to k_\mathrm{max}$ for hub activation, $\widetilde{\kappa}(\omega,\lambda) = k < k_\mathrm{max}$ for $\lambda > \lambda_c$. In words, for $\lambda$ beyond the absorbing phase, lower degree classes than $k_\mathrm{max}$ are able to self-sustain the dynamics in their neighborhood, largely increasing their infected density $\rho_k^*$. 

\begin{figure}
\includegraphics[width = 0.40\textwidth]{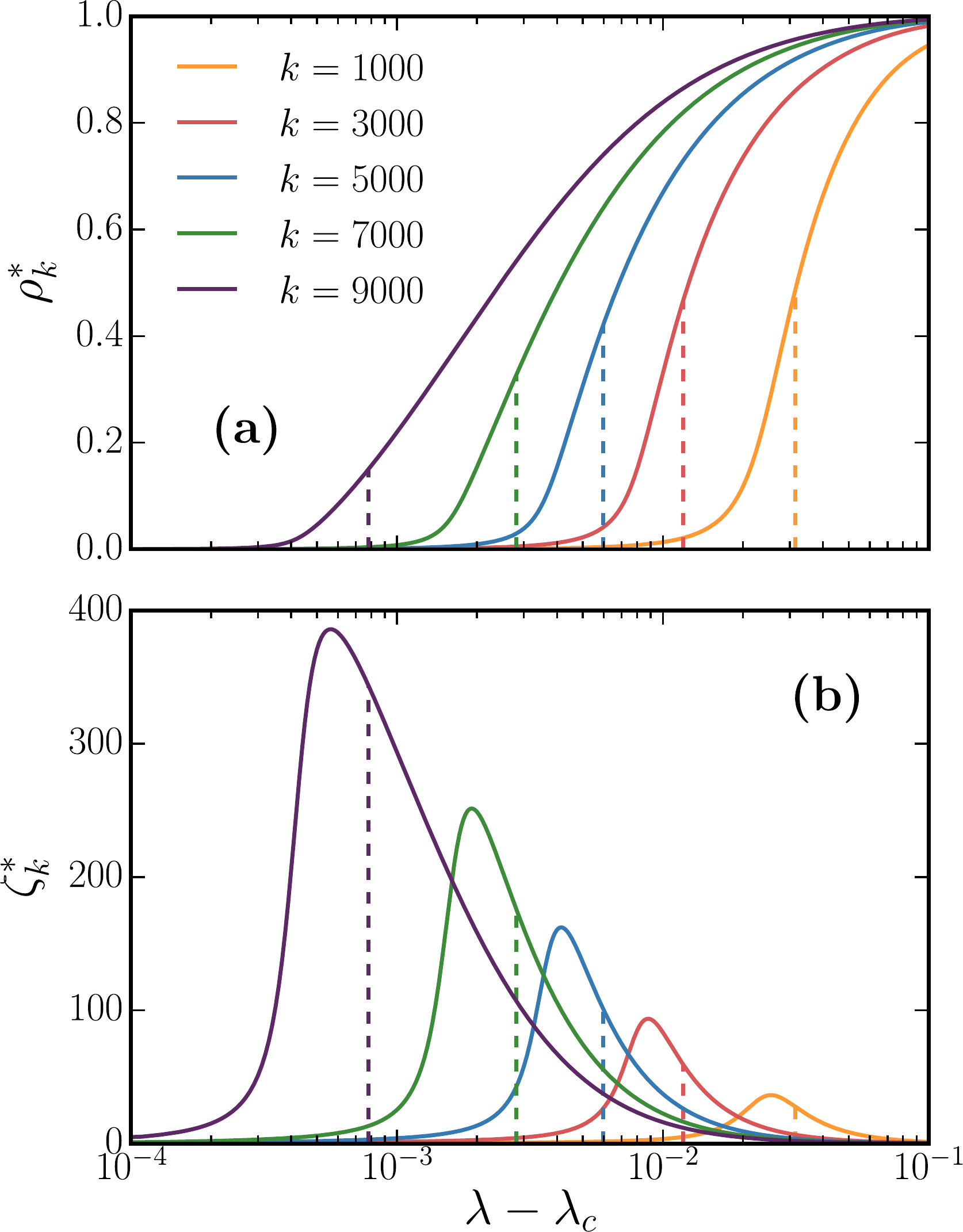}%
\caption{(Color online). Successive activation of the degree classes for a power-law degree distribution with exponent \mbox{$\gamma = 3.1$}, maximal degree $k_\mathrm{max} = 10^4$ and rewiring rate $\omega = 0$. The minimal degree is $k_\mathrm{min} = 3$. The vertical dashed lines corresponds to $\widetilde{\kappa} = k$ for each degree $k$ involved. (a) Infected density per degree class $\rho_k^*$ as a function of the infection rate $\lambda$. (b) Derivative of $\rho_k^*$ with respect to $\lambda$, $\zeta_k^*$, as a function of the infection rate $\lambda$. \label{fig:successive_activation}}
\end{figure}

This successive activation mechanism is observed in Fig.~\ref{fig:successive_activation}(a), where each $\rho_k^*$ sharply increases as $k \sim \widetilde{\kappa}$, then saturates according to Eq.~\eqref{rhoSS}. This is also well portrayed by the derivative of $\rho_k^*$ with respect to $\lambda$, $\partial_\lambda \rho_k^* \equiv \zeta_k^*$, which exhibits a maximum for $k \sim \widetilde{\kappa}$ [Fig.~\ref{fig:successive_activation}(b)]. 
\modified{These successive activations could be related to the smeared phase transition observed in Refs.~\cite{Odor2014,Cota2016} for power-law degree distribution with $\gamma > 3$. In a smeared phase transition, parts of the network exhibit an ordering transition independently, which in this case can be associated with the high degree nodes and their direct neighbors.}

\section{Conclusion \label{sec:conclusion}}

Using a degree-based theoretical framework, we have developed a stationary state analysis to study the SIS dynamics on time-varying configuration model networks. The rewiring mechanism has allowed us to take into account the effect of an effective structural dynamics, which mathematically represents an interpolation between a heterogeneous pair approximation (HPA) and a heterogeneous mean field theory (HMF). \modified{A general portrait of the phase transition that characterizes both collective and hub activation has emerged, filling the theoretical gap between degree-based and individual-based formalisms.}

\modified{First, we have shown that it is possible to discern the type of activation by studying the properties of $\theta_k^*$ near the absorbing phase, providing an alternative to the study of the principal eigenvector \cite{Goltsev2012}. This new point of view has inspired our analysis of the phase transition and allowed us to distinguish the hub and collective activation within our degree-based framework.}

\modified{
Second, by using a perturbative scheme, we have obtained a self-consistent expression for the absorbing-state threshold $\lambda_c$. Due to the analytical tractability of the RNA, we have been able to establish several correspondences with existing threshold expressions. Moreover, the generality of our threshold expression has allowed us to illustrate the impact of a time-varying structure by tuning the rewiring rate, leading to a smooth and possibly non-monotonic relation $\lambda_c(\omega)$. 
}

Third, by means of bounds on various quantities, we have characterized the critical exponents of $\rho^*$ and $\theta_k^*$ for power-law degree distributions. Noteworthy, it has allowed us to unveil the heterogeneous critical phenomenon for the hub activation scenario. \modified{This offers an elegant explanation for the heterogeneity of $\theta_k^*$ in Fig.~\ref{fig:neighborhood}(b) and also permits to discriminate between collective and hub-dominated phase transitions.}

Finally, we have studied the active phase beyond a hub activation threshold. \modified{The time variations of the structure leads to a more homogeneous neighborhood among the degree classes. Therefore, the dichotomy discussed in Sec.~\ref{subsec:collective_hub} is not as clear-cut anymore since the rewiring rate allows to interpolate between the two activation scenarios}. Also, in between the localized and delocalized regime for a hub-dominated phase transition, we have observed that each degree class undergoes a certain type of activation as the infection rate $\lambda$ is increased. \modified{These independent activations could be related to the smeared phase transition---with inhomogeneous ordering---observed in Refs.~\cite{Odor2014,Cota2016}.}

Several extensions of this work can be studied. For instance, the stationary state analysis can be applied to networks featuring other types of rewiring processes. These can be adaptive processes \cite{Gross2006,Gross2009,Marceau2010} or mechanisms that preserve other structural properties apart from the degree sequence, such as degree assortativity \cite{Newman2002}. Finally, due to the generality and versatility of the RNA, it can easily be applied to other binary-state dynamics.


\begin{acknowledgments}
We thank Laurent H\'ebert-Dufresne for useful discussions and comments. We acknowledge Calcul Qu\'{e}bec for computing facilities. This research was undertaken thanks to the financial support from the Natural Sciences and Engineering Research Council of Canada (NSERC), the Fonds de recherche du Qu\'{e}bec --- Nature et technologies (FRQNT) and the Canada First Research Excellence Fund.
\end{acknowledgments}

\appendix

\modified{\section{Development of the pair approximation \label{app:PA_derivation}}}

We adapt the approach proposed in Refs.~\cite{Marceau2010,Gleeson2011}, which starts with a set of differential equations governing the evolution of the compartments of nodes of a specified degree $k$ and infected degree $l$ (see also Refs.~\cite{Lindquist2011,Gleeson2013}). Let $s_{kl}(t)$ [$i_{kl}(t)$] be the probability that a degree $k$ node is susceptible (infected) and has $l \leq k$ infected neighbors. The rate equations for these probabilities are\begin{subequations}\label{ame}
\begin{align}
	\d{s_{kl}}{t} =& i_{kl} - \lambda l s_{kl}  + [1+\omega(1- \Theta)]\br{(l+1)s_{k(l+1)} - l s_{kl}}   \nonumber \\ +& (\Omega^S + \omega \Theta) \br{(k-l+1)s_{k(l-1)} - (k-l)s_{kl}} \;, \label{ame1}\\
	\d{i_{kl}}{t} =&  \lambda l s_{kl} - i_{kl} + [1+\omega(1- \Theta)]\br{(l+1)i_{k(l+1)} - l i_{kl} } \nonumber \\ +& (\Omega^I + \omega \Theta) \br{(k-l+1)i_{k(l-1)} - (k-l)i_{kl}} \;, \label{ame2}
\end{align}
\end{subequations}
where $\Omega^S(t)$ and $\Omega^I(t)$ are the mean infection rates for the neighbors of susceptible and infected nodes. These rates can be estimated from the compartmentalization \cite{Gleeson2011}, yielding
\begin{align}\label{defOmega}
	\Omega^S &= \lambda \frac{\sum_{l}\avg{(k-l)ls_{kl}} }{\sum_{l}\avg{(k-l)s_{kl}} } \;, & \Omega^I &= \lambda \frac{\sum_{l}\avg{l^2s_{kl}} }{\sum_{l} \avg{l s_{kl}} } \;.
\end{align}

Equations \eqref{ame} form an $\mathcal{O}\pr{k_\mathrm{max}^2}$ system of equations and do not lead to simple stationary solutions. To obtain a pair approximation formalism from Eqs.~\eqref{ame}, we use the dimensionality reduction scheme proposed in Ref.~\cite{Gleeson2011}. Let $\phi_k(t)$ be the probability of reaching an infected node following a random edge starting from a degree $k$ \textit{infected} node. Using Eqs.~\eqref{ame}, we can define a rate equation for $\theta_k$ and $\phi_k$ together with the definitions $\sum_l l s_{kl} = (1- \rho_k)k \theta_k$ and $\sum_l l i_{kl} = \rho_k k \phi_k$. This leads to the following system of equations
\begin{subequations}\label{eqThetaPhi}
\begin{align}
	\d{\theta_k}{t} =& -\frac{\lambda}{k(1- \rho_k)}\sum_l  l^2 s_{kl}  + r_k \phi_k + (\Omega^S + \omega \Theta) (1- \theta_k) \nonumber\\
	&-\br{1+\omega(1- \Theta)}\theta_k  - \theta_k \pr{ r_k - \lambda k \theta_k} \;, \label{eqTheta1}\\
	\d{\phi_k}{t} =& \frac{\lambda }{k \rho_k}\sum_l  l^2  s_{kl}  - \phi_k + (\Omega^I + \omega \Theta) (1- \phi_k) \nonumber\\
	&- \br{1+\omega(1- \Theta)}\phi_k  +  \phi_k  \pr{ 1 - \lambda k \theta_k r_k^{-1}} \;, \label{eqPhi1}
\end{align}
\end{subequations}
with $r_k \equiv \rho_k/(1-\rho_k)$. 

To obtain a closed system for Eqs.~\eqref{eqThetaPhi}, we use the \emph{pair approximation}
\begin{align}\label{2ndMoment}
 \sum_{l=0}^k l^2 s_{kl} \approx (1- \rho_k) \br{k \theta_k + k (k-1) {\theta_k}^2} \;,
\end{align}
which implies that the state of each neighbor is independent. The Eqs.~\eqref{eqThetaPhi_PA} and \eqref{defOmega2} follows accordingly.\\

\section{Monte-Carlo simulations \label{app:monte_carlo}}

To simulate the SIS dynamics on networks, we used a modified Gillespie algorithm \cite{Gillespie1976}. During the simulation process, we track the number of infected nodes $n(t)$ and the number of stubs emanating from them $u(t)$. The total number of stubs is $2M$ and is fixed according to our rewiring process. At each step, three event types are possible with the following probability
\begin{subequations}
\begin{align}
	P(\mathrm{Recovery}) &= n/(n + \lambda u + \omega M/2) \; , \\
	P(\mathrm{Infection}) &= \lambda u/(n + \lambda u + \omega M/2) \;, \\
	P(\mathrm{Rewiring}) &= (\omega M/2)/(n + \lambda u + \omega M/2) \;.
\end{align}
\end{subequations}
Each event occurs as follows
\begin{itemize}
	\item Recovery event : an infected node is chosen randomly and becomes susceptible.
	\item Infection attempt event : an infected node is chosen proportionally to its degree. We then choose one of its emanating stubs randomly and infect the node at the other end point. If it is already infected, we do nothing : this phantom process \cite{Cota2017} corrects the probability in order to make the process equivalent to randomly choosing an edge among the set of all susceptible-infected edges.
	\item Rewiring event : Two edges $(a_1, b_1)$ and $(a_2, b_2)$ are randomly chosen with $a_i, b_i$ the labels for the nodes; choosing an edge $(b_1, a_1)$ is equally likely. We then rematch the stubs according to the following scheme $(a_1, b_1), (a_2, b_2) \mapsto (a_1, b_2), (a_2, b_1)$. Loops and multi-edges are permitted.
\end{itemize}
After all events---even the frustrated ones---we update the time with $t \mapsto t + \Delta t$ where $\Delta t \equiv E\br{\Delta t} = [n(t) + \lambda u(t) + \omega M/2]^{-1}$.

To evaluate some observables for infection rates $\lambda$ near the absorbing phase, we sample the configurations of the system that do \emph{not} fall on the absorbing state---the quasi-stationary distribution \cite{Marro2005,Oliveira2005,Ferreira2011,Sander2016}. When the system visits the absorbing state, the current state is replaced by a configuration randomly chosen among the set $\mathcal{H}$ of previously stored active configurations. Also, with probability $\xi \Delta t$, each active configuration is stored, replacing a randomly chosen one among $\mathcal{H}$, thus updating the set of states proportionally to their average lifetime \cite{Oliveira2005}. The system is then expected to converge on the quasi-stationary distribution \cite{Blanchet2014} over which we measure observables. In all our simulations, we chose $|\mathcal{H}| \in [50,100]$ and $\xi = 10^{-2}$. \\

\section{Supplementary developments for the critical exponents \label{app:SM_critical_exponent}}

\subsection{Lower and upper bounds on {\boldmath$\theta_k^*$} \label{app:lower_upper_bound}}

Our insight is that $\theta_k^*$ is a monotically increasing function of the degree $k$. Higher degree nodes have a higher probability of being infected, hence their neighbors can only be more infected on average. This is reflected in Eq.~\eqref{thetaSol}, despite not being explicit.

The lower and upper bounds are then fixed using the minimal and maximal values for the degree in Eq.~\eqref{thetaSol}.
\begin{align}
\theta_{-}^* &\equiv \br{\frac{\beta}{\kappa}}_{-} \leq \frac{\beta}{\kappa - 1} \;, \\
\theta_{+}^* &\equiv \br{\frac{1}{\alpha}}_{+} = \lim_{k\to \infty} \theta_k^* \;.
\end{align}
The parameters $\alpha,\beta,\kappa$ are considered finite when taking the limit $k \to \infty$ in the second equation, which is true for any $\lambda > \lambda_c$. 

\subsection{Integral approximation \label{app:integral_evaluation}}

Let us consider an integral of the form
\begin{align}
	I = {k'}^{-a}b^{-1} \int_{k'}^{\infty} \frac{k^{a-1} }{1 + k(bk')^{-1}}\dx k \;,
\end{align}
where $b \equiv (\lambda \theta^* k')^{-1}$ and $a <1$, equal to $(3-\gamma)$ or $(2-\gamma)$ according to the integrals appearing in Eq.~\eqref{integralFormOmegaMin}. Using $z \equiv k' k^{-1}$, this can be rewritten as
\begin{align}
	I = \int_0^1 \frac{z^{-a}}{1+bz} \dx z \;.
\end{align}
This integral can be associated with the hypergeometric function \cite{Gradshteyn2014}
\begin{align}\label{integral_2F1_rep}
	I = (1-a)^{-1} {}_2F_1(1,1-a;2-a;-b) \;.
\end{align}
Since near the absorbing phase $b \gg 1$, to extract the leading terms of Eq.~\eqref{integral_2F1_rep}, we use the transformation formulas for the hypergeometric function \cite{Gradshteyn2014}, leading to
\begin{align}
	I = \Gamma(1-a)\Gamma(a) b^{a-1} - (ab)^{-1} {}_2F_1\pr{1,a;a+1;-b^{-1}} \;.
\end{align}
The leading terms are finally 
\begin{align}
	I = h_1 b^{a-1} + h_2 b^{-1} + \mathcal{O}\pr{b^{-2}} \;,
\end{align}
where $\cbrace{h_i}$ are non-vanishing constants. Appropriate limits must be taken for all values of $a = 0$  or negative integer values.  

\subsection{Critical behavior of {\boldmath$\theta_{k_\mathrm{min}}^*$} and {\boldmath$\theta_{k_\mathrm{max}}^*$} \label{app:theta_critical_behavior}}

Near the phase transition ($\lambda \to 0$ in this case), according to Eq.~\eqref{kappaDef}, $\kappa \simeq \widetilde{\kappa}(\omega,\lambda)$ is very large. Since we can choose $\lambda$ arbitrarily small, we can let $\kappa \to \infty$, keeping however $\kappa \ll k_\mathrm{max} \to \infty$.

For $\theta_{k_\mathrm{min}}^*$, we simply use the perturbative development [Eq.~\eqref{thetaDev}] to extract the leading term
\begin{align}
	\theta_{k_\mathrm{min}}^* &= \frac{\beta}{\kappa - k_\mathrm{min}} + \mathcal{O}(\beta^2) \simeq \frac{\beta}{\kappa} \;.
\end{align}
For $\theta_{k_\mathrm{max}}^*$, we need to develop Eq.~\eqref{thetaSol} in terms of $\kappa/k_\mathrm{max} \to 0$ instead. In this case, we obtain
\begin{align}
	\theta_{k_\mathrm{max}}^* = \frac{1}{\alpha} + \mathcal{O}\pr{\frac{\kappa}{k_\mathrm{max}}} \simeq \frac{1}{\alpha}\;.
\end{align}


%

\end{document}